\documentclass[final, journal]{IEEEtran}
\usepackage{cite}

\usepackage[printonlyused]{acronym} % Acronyms

\usepackage[cmex10]{amsmath} % cmex10 option prevents amsmath from using a Type 3 font for math within footnotes.
\usepackage{amsthm} % AMS:Theorems
\usepackage{amssymb} % AMS:Symbols
\usepackage{array}
\usepackage{cancel} % Strikeout part of the equation (\cancel, \bcancel, \xcancel)

\usepackage[caption=false,font=footnotesize]{subfig}
\usepackage[pdftex]{graphicx}
\graphicspath{{./figs/}}
\DeclareGraphicsExtensions{.pdf,.jpeg,.png}

\usepackage{color}

\usepackage{balance} % Balanced two-column mode
\usepackage{bm} % Makes the given argument bold

\usepackage{threeparttable} % Table
\usepackage{multicol} % Table: multicolumn
\usepackage{multirow} % Table: multirow
\usepackage{rotating} % Table: rotate text

\usepackage{upgreek} % upgreek greek letters
\usepackage{url} % Verbatim with URL-sensitive line breaks
\usepackage{verbatim} % Preformatted text blocks

\usepackage{mathtools} % Commands: coloneqq eqqcolon
\usepackage{mathrsfs} % Provides a \mathscr command

\usepackage{xfrac} % Split-level fractions

\usepackage{fancyhdr}
\usepackage{amsmath}

\usepackage{etoolbox}

\newcommand{\herm}{^{\mbox{\scriptsize H}}}

\newcommand{\Starherm}{^{\mbox{\scriptsize *H}}}

\newcommand{\tran}{^{\mbox{\scriptsize T}}}

\newcommand{\comb}{^{\mbox{\scriptsize c}}}

\newcommand{\cherm}{^{\mbox{\scriptsize cH}}}

\newcommand{\iLoop}{^{{\scriptsize (i)}}}

\newcommand{\xiLoop}{^{{\scriptsize (i-1)}}}

\newcommand{\iherm}{^{\scriptsize(i){\mbox{\scriptsize H}}}}
\newcommand{\xiherm}{^{\scriptsize(i-1){\mbox{\scriptsize H}}}}

\newcommand{\HypTst}[2]{\raise2.5ex\hbox{\scriptsize$#1$} \hspace{-1.3em}\displaystyle\gtreqless\hspace{-1.2em}\raise-1.1em\hbox{\scriptsize$#2$}}

\newcommand{\cref}[1]{C\ref{#1}}
\acrodef{3-G}{3-Generation}
\acrodef{3GPP}{3rd Generation Partnership Project}
\acrodef{a.k.a}{also-known-as}
\acrodef{AAS}{active antenna system}
\acrodef{ADMM}{alternating direction method of multipliers}
\acrodef{AoA}{angle-of-arrival}
\acrodef{AoD}{angle-of-departure}
\acrodef{AHB}{Abel hybrid bound}
\acrodef{A-TS}{alternating Taylor\'s series}
\acrodef{A-LASSO}{\emph{adaptive}-least absolute shrinkage and selection operator}
\acrodef{AWGN}{Additive White Gaussian Noise} 
\acrodef{AGC}{Automated Gain Control}
\acrodef{BER}{Bit Error Rate}
\acrodef{BB}{Bhattacharyya Bound}
\acrodef{BF}{beamformer}
\acrodef{BFGS}{Broyden-Fletcher-Goldfarb-Shanno}
\acrodef{BS}{base-station}
\acrodef{BBU}{baseband processing unit}
\acrodef{cdf}{cumulative distribution function}
\acrodef{CEP}{Circular Error Probability}
\acrodef{CoMP}{coordinated multi-point}
\acrodef{CIR}{Channel Impulse Response}
\acrodef{C-RDM}{Cross-Ranging Direction Matrix}
\acrodef{C-NLS}{Constrained Non-linear Least Squares}
\acrodef{CRLB}{Cram\'{e}r-Rao Lower Bound}
\acrodef{CSIT}{channel state information at the transmitter}
\acrodef{CSI}{channel state information}
\acrodef{CP}{Cyclic-Prefix}
\acrodef{CB}{coordinated beamforming}

\acrodef{DC}{Distance Contraction}
\acrodef{DCT}{Distance Contraction Theory}
\acrodef{DE}{Distance Error}
\acrodef{DFT}{Discrete Fourier Transform}
\acrodef{DL}{Downlink}
\acrodef{DoA}{Direction-of-Arrival}
\acrodef{DSSS}{Direct Sequence Spread Spectrum}
\acrodef{DoF}{degree-of-freedom}
\acrodef{DPS}{dynamic point selection}

\acrodef{EDF}{Euclidean Distance Function}
\acrodef{EDM}{Euclidean Distance Matrix}
\acrodef{EFIM}{Equivalent Fisher Information Matrix}
\acrodef{EHF}{Extremely High Frequency}
\acrodef{EKT}{Euclidean Kernel Transformation}
\acrodef{EK}{Euclidean Kernel}
\acrodef{ERII}{Equivalent Ranging Information Intensity}
\acrodef{ETSI}{European Telecommunications Standards Institute}
\acrodef{FIM}{Fisher Information Matrix}
\acrodef{FCC}{Federal Communications Commission}
\acrodef{FDD}{Frequency-Division-Duplex}
\acrodef{GLE}{Geometric-constrained Location Estimation}
\acrodef{G-LS}{Global Least Squares}
\acrodef{GNSS}{Global Navigation Satellite System}
\acrodef{GP}{Gaussian Process}
\acrodef{GP-LVM}{Gaussian Process Latent Variable Model}
\acrodef{GPS}{Global Positioning System}
\acrodef{GDC}{Global Distance Continuation}
\acrodef{GDOP}{Geometric Dilution of Precision}
\acrodef{GTRS}{Generalized Trust Region Subproblems}
\acrodef{HCRB}{Hammersley-Chapmann-Robbins Bound}
\acrodef{HB}{Hybrid Bound}
\acrodef{HOSVD}{High-Order Singular-Value-Decomposition}
\acrodef{HPBW}{half power beamwidth}
\acrodef{iid}{independent identically distributed}
\acrodef{ICT}{Information Communication Technology}
\acrodef{I-EKT}{Inverse Euclidean Kernel Transformation}
\acrodef{IoT}{Internet-of-Things}
\acrodef{IIoT}{Industrial Internet-of-Things}
\acrodef{iff}{if and only if}
\acrodef{IFT}{Inverse Fourier Transform}
\acrodef{ITU}{International Telecommunication Union}
\acrodef{I/N}{Interference-to-Noise}
\acrodef{JT}{joint transmission}
\acrodef{JT-CoMP}{joint transmission coordinated multi-point}

\acrodef{KKT}{Karush-Kuhn-Tucker}
\acrodef{KLT}{Karhunen-Lo\'{e}ve Transform}
\acrodef{LASSO}{least absolute shrinkage and selection operator}
\acrodef{LARSO}{Least Absolute Residual and Selection Operator}
\acrodef{LBSs}{Location Based Services}
\acrodef{LM}{Levenberg-Marquardt}
\acrodef{LQ}{Link-Quality}
\acrodef{LS}{Least Squares}
\acrodef{LLS}{Linearized Least Squares}
\acrodef{LT}{Location-Tracking}
\acrodef{LTE}{Long Term Evolution}
\acrodef{LoS}{line-of-sight}
\acrodef{LOESS}{Locally weighted Scatterplot Smoothing}
\acrodef{LP}{Linear Programming}
\acrodef{MAP}{Maximum A Posteriori}
\acrodef{MOO}{Multiple-Objective Optimization}
\acrodef{MOS}{Mean Opinion Score}
\acrodef{MIMO}{multiple-input-multiple-output}
\acrodef{MU-MIMO}{multi-user multiple-input-multiple-output}
\acrodef{MU-MISO}{multi-user multiple-input-single-output}
\acrodef{MMSE}{Minimum Mean Squared Error}
\acrodef{mmWave}{millimeter-wave}
\acrodef{MSE}{Mean Square Error}
\acrodef{ML}{Maximum Likelihood}
\acrodef{MS}{mobile-station}
\acrodef{MDS}{Multidimensional Scaling}
\acrodef{MSPE}{Mean-Squared-Position-Error}
\acrodef{MUSIC}{MUltiple SIgnal Classification}
\acrodef{MRT}{maximum ratio transmission}
\acrodef{MURM}{Minimum User-Rate Maximization}

\acrodef{N-DC}{Negative-Distance Contraction}
\acrodef{NLOS}{non-line-of-sight}
\acrodef{NLS}{Non-Linear-Least Squares}
\acrodef{NP-hard}{Non-deterministic Polynomial-time hard}
\acrodef{NPRM}{Notice of Proposed Rulemaking}
\acrodef{OFDM}{orthogonal frequency-division multiplexing}
\acrodef{OMP}{Orthogonal Matching Pursuit}
\acrodef{OTDoA}{observed-Time-Difference-of-Arrival}
\acrodef{psd}{positive semi-definite}
\acrodef{PSD}{power spectral density}
\acrodef{pdf}{probability density function}
\acrodef{PDP}{Power Delay Profile}
\acrodef{P-DC}{Positive-Distance Contraction}
\acrodef{PE}{Position Error}
\acrodef{PEB}{Position Error Bound}
\acrodef{PREB}{position-rotation error bound}
\acrodef{PSM}{Positive Semi-definite Matrix}
\acrodef{POCS}{Projection On Convex Sets}
\acrodef{PCA}{Principal Component Analysis}
\acrodef{PPCA}{Probabilistic Principal Component Analysis}
\acrodef{QoD}{Quality-of-Design}
\acrodef{QoL}{Quality-of-Location}
\acrodef{QoS}{Quality-of-Service}
\acrodef{QoE}{Quality-of-Experience}
\acrodef{QoI}{Quality-of-Information}
\acrodef{R95}{95$\%$ Radius}
\acrodef{RBF}{Radial Basis Function}
\acrodef{RDM}{Ranging Direction Matrix}
\acrodef{REB}{Rotation Error Bound}
\acrodef{RF}{radio frequency}
\acrodef{R-GDC}{Range-Global Distance Continuation}
\acrodef{R-LS}{Regularized Least-Squares}
\acrodef{RSS}{Received Signal Strength}
\acrodef{RSSI}{Received Signal Strength Index}
\acrodef{RMB}{Reuven-Messer Bound}
\acrodef{RMSE}{root-mean-squared-error}
\acrodef{RII}{Ranging Information Intensity}
\acrodef{RR}{Ridge-Regression}
\acrodef{RRU}{remote radio unit}

\acrodef{SCE}{Sparse Channel Estimator}
\acrodef{SCM}{spatial channel model}
\acrodef{SDP}{Semi Definite Programming}
\acrodef{SIMO}{Single-Input-Multiple-Output}
\acrodef{SISO}{Single-Input-Single-Output}
\acrodef{SLAM}{Simultaneous Localization and Mapping}
\acrodef{SMACOF}{Stress-of-a-MAjorizing-Complex-Objective-Function}
\acrodef{SQP}{Sequential Quadratic Programming}
\acrodef{SR-LS}{Squared-Range Least-Squares}
\acrodef{SR-GDC}{Square Range-Global Distance Continuation}
\acrodef{SER}{Symbol-Error-Rate}
\acrodef{SB}{Stochastic Bound}
\acrodef{SNR}{Signal-to-Noise Ratio}
\acrodef{SHF}{Super High Frequency}
\acrodef{SINR}{signal-to-interference-noise ratio}
\acrodef{SD}{Steepest-Descent}
\acrodef{SA}{Simulated-Annealing}
\acrodef{SR}{Squared Range}
\acrodef{SCA}{successive convex approximation}
\acrodef{SOCP}{second-order cone program}
\acrodef{TDoA}{Time-Difference-of-Arrival}
\acrodef{TS}{Taylor's Series}
\acrodef{ToF}{Time-of-Flight}
\acrodef{ToA}{Time-of-Arrival}
\acrodef{TDD}{time division duplex}
\acrodef{TTI}{Transmission Time Interval}
\acrodef{ULA}{uniform linear array}
\acrodef{UMTS}{Universal Mobile Telecommunications System}
\acrodef{URA}{uniform rectangular array}
\acrodef{UWB}{Ultra-WideBand}
\acrodef{UHF}{Ultra High Frequency}
\acrodef{WLS}{Weighted Least Squares}
\acrodef{WC}{Weighted Centroid}
\acrodef{WSRM}{weighted sum-rate maximization}
\acrodef{ZC}{Zadoff-Chu}
\acrodef{ZF}{Zero-Forcing}

\IEEEoverridecommandlockouts

\def\@IEEEfigurecaptionsepspace{\vskip\abovecaptionskip\relax}%

\usepackage{optidef}
\usepackage[ruled,linesnumbered]{algorithm2e}
\newcounter{mytempeqncnt}

\usepackage{bbm}

\newcommand{\revise}[1]{{\color{black} #1}}
\begin{document}
\title{Blockage-aware Reliable mmWave Access via Coordinated Multi-point Connectivity}
%
%%\author{
%%\IEEEauthorblockN{Dileep Kumar, Jarkko Kaleva, Antti T\"{o}lli}
%%\IEEEauthorblockA{\textit{Centre for Wireless Communications, 
%%University of Oulu, Oulu, Finland} \\
%%Email: \{dileep.kumar, antti.tolli\}@oulu.fi, jarkko.kaleva@solmutech.com}
%%	\thanks{This work was supported by Academy of Finland under grants~no.~313041 (PRISMA: Positioning-aided Reliably-connected Industrial Systems with Mobile mmWave Access) and 318927 (6Genesis Flagship).}
%%}
\author{
\IEEEauthorblockN{Dileep~Kumar,~\IEEEmembership{Student~Member,~IEEE,} Jarkko~Kaleva,~\IEEEmembership{Member,~IEEE,} and~Antti~T\"{o}lli~\IEEEmembership{Senior~Member,~IEEE}}
\thanks{This work was supported in parts by the European Commission in the framework of the H2020-EUJ-02-2018 project under grant~no.~815056 ({5G}-Enhance) and in parts by Academy of Finland under grants~no.~313041 (PRISMA: Positioning-aided Reliably-connected Industrial Systems with Mobile mmWave Access), 311741 ({WiFiUS}) and 318927 (6Genesis Flagship). The work of D.~Kumar was supported in part by Nokia Foundation, and in part by Riitta ja Jorma J. Takanen Foundation, and in part by Tauno T{\"o}nningin S{\"a}{\"a}ti{\"o}n Foundation.
}% <-this % stops a space
\thanks{
 This article was presented in parts at the IEEE 27th European Signal Processing Conference, A Coru\~{n}a, Spain, Sep 2019, and in parts at the IEEE Global Communications Conference, Waikoloa, HI, USA, Dec 2019. \textit{(Corresponding~author: Dileep~Kumar)}}
\thanks{Authors Dileep Kumar and Antti T\"olli are with Centre for Wireless Communications, University of Oulu, FIN-90014 Oulu, Finland. (e-mail: dileep.kumar@oulu.fi;  antti.tolli@oulu.fi). Jarkko Kaleva is with Solmu Technologies (e-mail: jarkko.kaleva@solmutech.com).}% <-this % stops a space
}

%%%header-Begin
\makeatletter
\let\old@ps@headings\ps@headings
\let\old@ps@IEEEtitlepagestyle\ps@IEEEtitlepagestyle
\def\confheader#1{%
% for all pages except the first
%%\def\ps@headings{%
%%\old@ps@headings%
%%\def\@oddhead{\strut\hfill#1\hfill\strut}%
%%\def\@evenhead{\strut\hfill#1\hfill\strut}%
%%}%
% for the first page
\def\ps@IEEEtitlepagestyle{%
\old@ps@IEEEtitlepagestyle%
\def\@oddhead{\strut\hfill#1\hfill\strut}%
\def\@evenhead{\strut\hfill#1\hfill\strut}%
}%
\ps@headings%
}
\makeatother

\confheader{%
\indent \footnotesize{Accepted for publication in IEEE Transactions on Wireless Communications in Jan. 2021. (Digital Object Identifier: 10.1109/TWC.2021.3057227)}  
}
%%%header-Ends

\maketitle

%% ABSTRACT -----------------------------------------
\begin{abstract}
%%%%%%%%%%%%%%%%%%%%%%%%%%%%%  200 WORD LIMITS   %%%%%%%%%%%%%%%%%%%%%%%%%%%%%%%%%%%%%%%%%%
%%%%%%%%%%%%%%%%%%%%%%%%%%%%%  READ Following ONLY   %%%%%%%%%%%%%%%%%%%%%%%%%%%%%%%%%%%%%%%%
%
The fundamental challenge of the millimeter-wave ({mmWave}) frequency band is the sensitivity of the radio channel to blockages, which gives rise to unstable connectivity and impacts the reliability of a system. To this end, multi-point connectivity is a promising approach for ensuring the desired rate and reliability requirements.    
%
%%%%A robust beamformer design is proposed in this paper for improved communication reliability by exploiting multi-point connectivity and a pessimistic estimate of achievable rates over potential link blockage combinations. 
%
A robust beamformer design is proposed to improve the communication reliability by exploiting the spatial macro-diversity and a pessimistic estimate of rates over potential link blockage combinations.  
Specifically, we provide a  blockage-aware algorithm for the weighted sum-rate maximization ({WSRM}) problem with parallel beamformer processing across distributed remote radio units ({RRUs}). Combinations of non-convex and coupled constraints are handled via successive convex approximation ({SCA}) framework, which admits a closed-form solution {for each {SCA} step}, by solving a system of Karush-Kuhn-Tucker ({KKT}) optimality conditions. 
Unlike the conventional coordinated multi-point ({CoMP}) schemes, the proposed blockage-aware beamformer design has, per-iteration, computational complexity in the order of {RRU} antennas instead of system-wide joint transmit antennas. This leads to a practical and computationally efficient implementation that is scalable to any arbitrary multi-point configuration.   %%%%, as well as, for any desired rate-reliability trade-off.   
In the presence of random blockages, the proposed schemes are shown to significantly outperform baseline scenarios and result in reliable {mmWave} communication.
%
%%%%%%%%%%%%%%%%%%%%%%%%%%%%%%%%%%%%%%%%%%%%%%%%%%%%%%%%%%%%%%%%%%%%%%%%%%%%%%%%%%%%%%%%%%%%%%%%%%%%%%
%
%
\end{abstract}


%%%%\vspace{-5px}
%% INTRODUCTION -----------------------------------------
\section{Introduction}
The proliferation of ever-increasing data-intensive wireless applications along with spectrum shortage motivates the investigation of \ac{mmWave} communication for the upcoming $5$th-generation ($5$G) New Radio (NR) and beyond cellular systems~\cite{Rappaport-MillimeterWave-2013, Andrews-WhatWill-2014}. The \ac{mmWave} frequency band not only provides relatively large system bandwidth but also allows for packing a significant number of antenna elements for highly directional communication~\cite{Rappaport-MillimeterWave-2013}, which is important to ensure link availability as well as to control interference in dense deployments~\cite{Andrews-WhatWill-2014}. 
Hence, the \ac{mmWave} mobile communication is anticipated to substantially increase the average system throughput. %
%%%There are still many issues that need to be resolved before these technologies are ready for the commercial applications. 
%
However, the fundamental challenge is the sensitivity of \ac{mmWave} radio channel to blockages due to reduced diffraction, higher path and penetration loss~\cite{MacCartney-RapidFading-2017,  AndreevBlockage2018}. These lead to rapid degradation of signal strength and give rise to unstable and unreliable connectivity. For example, a mobile human blocker can obstruct the dominant paths for hundreds of {milliseconds}, and normally lead to disconnecting the communication session~\cite{MacCartney-RapidFading-2017, AndreevBlockage2018}. On the other hand, finding an alternate unblocked direction causes critical latency overheads.
%%%%Furthermore, connection reliability is aggravated by the fact that, for instance, a human blocker can degrade the channel quality by $20-30$~dB for up to hundreds of millisecond~\cite{MacCartney-RapidFading-2017}. 
Hence, the presence of such frequent and long duration blockages significantly reduces the experienced quality-of-service (QoS)~\cite{AndreevBlockage2018}. 
To overcome such challenges, use of \ac{CoMP} schemes, where the users are concurrently connected to multiple \acp{RRU}, are highly useful for providing more robust and resilient  communication~\cite{Antti-OntheValue-2009, Antti-CooperativeMIMO-OFDM-2008, Wang-CoMP-2016,  irmer-coordinated-2011, Nigam-CoMP,  MacCartney-BaseStationDiversity-2017, MacCartney-BSDiversityJournal-2019, Maamari-CoverageinmmWave-2016, Andreev-Multiconnectivity-2019, Skouroumounis-Low-Complexity-Journal-2017, Lee-Hybrid_2017}. Therefore, it is envisioned that multi-connectivity schemes by utilizing the multi-antenna spatial redundancy via geographically separated transceivers will be of high importance in future mmWave systems~\cite{NR3GPP-MultiConnectivity}. %%%%%%%Moreover, Qualcomm has recently implemented a 5G-CoMP testbed~\cite{Qualcomm} in order to improve coverage, capacity and reliability performance through a flexible deployment scenarios. 

%\vspace{-10px}
%% ------- Prior Work -----------------------------------------
\subsection{Prior Work}
\label{subsec:PriorWork}
The \ac{CoMP} transmission and reception are typically used to increase the system throughput, particularly for the cell-edge users due to relatively long distance from the serving \ac{BS} and adverse channel conditions (e.g., higher path-loss and interference from neighboring BSs). Such scenarios have been widely studied over the past decade {in the context of}  $4$th-generation ($4$G)~systems~\cite{Antti-OntheValue-2009, Antti-CooperativeMIMO-OFDM-2008, irmer-coordinated-2011, Nigam-CoMP, Wang-CoMP-2016}. Techniques, such as,  \ac{JT}, \ac{CB} and \ac{DPS} were standardized in \ac{3GPP} and were widely studied in Long Term Evolution-Advanced (LTE-A) to enhance capacity and converge by efficiently utilizing the spatially separated transceivers~\cite{irmer-coordinated-2011}. For example, it is shown in~\cite{Nigam-CoMP} that \ac{JT}-\ac{CoMP} increases the coverage by, up to, $17\%$ for general users and $24\%$ for cell-edge users compared to non-cooperative~scenario.

Recent studies have considered the deployment of \ac{CoMP} in the mmWave frequencies~\cite{MacCartney-BaseStationDiversity-2017, MacCartney-BSDiversityJournal-2019, Maamari-CoverageinmmWave-2016, Andreev-Multiconnectivity-2019, Skouroumounis-Low-Complexity-Journal-2017, Lee-Hybrid_2017}. Also, it is considered in \ac{3GPP} for upcoming $5$G NR and beyond mmWave based cellular systems~\cite{NR3GPP-MultiConnectivity}.
In~\cite{MacCartney-BaseStationDiversity-2017, MacCartney-BSDiversityJournal-2019}, the authors showed a  significant coverage improvement by simultaneously serving a user with spatially distributed transmitters. Results were drawn from extensive real-time measurements for $73$~GHz in the urban open square scenario. The network coverage gain for the \ac{mmWave} system with multi-point connectivity, in the presence of random blockages, was also confirmed in~\cite{Maamari-CoverageinmmWave-2016, Andreev-Multiconnectivity-2019} using stochastic geometry tools. The work in~\cite{Skouroumounis-Low-Complexity-Journal-2017} proposed a low complexity cooperation technique for the \ac{JT}, wherein a subset of cooperating \acp{BS} is obtained by selecting the strongest BS in each tier. The authors also investigated the impact of blockage density in heterogeneous multi-tier network. 
Similarly to earlier works on single-cell two-stage hybrid analog-digital beamforming design, e.g., in~\cite{Alkhateeb-LimitedFeedback-2015,Yu-Alternating-2016}, authors in~\cite{Lee-Hybrid_2017} considered a multi-user massive multiple-input-multiple-output~(MU-MIMO) system with \ac{JT}-\ac{CoMP} processing where a high-dimensional analog beamformer is followed by a low-dimensional centralized digital baseband precoder.
However,~\ac{CoMP} techniques in~\cite{Antti-OntheValue-2009, Antti-CooperativeMIMO-OFDM-2008, irmer-coordinated-2011, Nigam-CoMP, Wang-CoMP-2016, MacCartney-BaseStationDiversity-2017, MacCartney-BSDiversityJournal-2019, Maamari-CoverageinmmWave-2016, Andreev-Multiconnectivity-2019, Skouroumounis-Low-Complexity-Journal-2017, Lee-Hybrid_2017} were still devised with the sole scope of enhancing the capacity and coverage by utilizing the spatially separated transceivers. Thus, they were not originally designed for the stringent reliability requirements of, e.g., industrial-grade critical applications.

It is well known that a system can provide any level of reliability by sequential data transmission, i.e., by retransmitting the same message at various protocol levels, until a receiver acknowledges correct reception over a dedicated feedback channel~\cite{Johansson-Radioaccess-2015}. However, in the presence of random link blockages, high penetration and path-loss, \ac{mmWave} feedback links are inherently unreliable and, hence, they require  redundant retransmissions. On the other hand, allowable latency dictates a strict upper limit on the number of retransmissions~\cite{Shariatmadari-Linkadaptation-2016}.%%%% attempts~\cite{Shariatmadari-Linkadaptation-2016}.

The loss of connection in the \ac{mmWave} communication is mainly due to a sudden blockage of the dominant links, generally caused by abrupt mobility, self-blockage or external blockers~\cite{MacCartney-RapidFading-2017, AndreevBlockage2018}. Accurate estimation of each blocker requires precise environment mapping and frequent \ac{CSI} acquisition, which might result in significant coordination overhead.   %%% and severe synchronization challenges. 
Furthermore, blockage events can create large latencies if a passive hand-off is inevitable~\cite{Shariatmadari-Linkadaptation-2016}. 
Thus, the limitations of retransmission events and the difficulty of accurate estimation of random blocking events motivate us to develop more robust and resilient downlink transmission strategies that can retain stable connectivity under the uncertainties of \ac{mmWave} channels and random blockages.     

%%%%\vspace{-10px}
%% ------- Contributions -----------------------------------------
\subsection{Contributions}
\label{subsec:Contribution}

Motivated by the above concerns, we propose a robust beamforming design for the \ac{JT}-\ac{CoMP}, which improves the sum-rate while retaining stable and resilient connectivity for mmWave mobile access in the presence of random blockers. The key contributions of this paper include:  
\begin{itemize}
    \item A blockage-aware beamformer design with a strong emphasis on system reliability is provided by exploiting multi-antenna spatial diversity and \ac{CoMP} connectivity. The weighted downlink sum-rate is maximized\footnote{The formulation can be easily modified to handle other objective functions, {e.g., equal rate allocation or queue minimization.}}, where, for each user, a pessimistic estimate of the achievable rate over all possible combinations of potentially blocked links among the cooperating \acp{RRU} is considered. Managing a large set of link blockage combinations is considerably more difficult than conventional  constrained optimization~\cite{Wang-CoMP-2016, Lee-Hybrid_2017, Kaleva-DecentralizeSumRate-2016, Ganesh-TrafficAware-2016, Tervo-EE-2017, Kaleva-DecentralizeJP-2018, Tervo-EE-2018} due to the mutually coupled \ac{SINR} constraints.  
     The preemptive modeling of serving set over the potential link blockage combinations are shown to greatly improve the system outage performance while ensuring user-specific rate and reliability requirements.
     %
     %%%The proposed schemes are shown to greatly improve the outage performance while ensuring any desired rate-reliability requirements by preemptive modeling of serving set over the potential link blockage combinations.    
   %%%%with the proposed scheme  --- can achieve any desired rate - reliability targets... based on e.g. some application specific requirements. analysis is provided...  cover eqs (8-9) and Appendix B here...
%
    \item A  \ac{SCA} based beamforming algorithm is provided for the original non-convex and computationally challenging problem. More specifically, all coupled and non-convex constraints are conservatively approximated with a sequence of convex subsets and iteratively solved until convergence. The underlying subproblems, for each \ac{SCA} iteration, become \acp{SOCP}, and that are efficiently solvable by any standard off-the-shelf solvers.
    \item A  low-complexity robust beamformer design framework is proposed that merges the \ac{SCA} with dual~\cite{boyd2003subgradient} and best response~\cite{Kaleva-DecentralizeJP-2018} methods to admit parallel beamformer processing for the distributed \acp{RRU} via iterative evaluation of the closed-form \ac{KKT} optimality conditions. The schemes proposed in~\cite{Kaleva-DecentralizeSumRate-2016, Kaleva-DecentralizeJP-2018} cannot be used directly, thus our proposed \ac{KKT} based solution is significantly more advanced, and provides an approach for solving mutually coupled minimum \ac{SINR} constraints. 
    This leads to a practical, latency-conscious, and computationally efficient implementation for cloud edge architecture.
    \item  A detailed implementation of proposed methods is provided assuming digital beamforming architecture. Moreover, for completeness, a low-complexity two-stage hybrid analog-digital beamforming implementation is introduced in the numerical section. As a result, the  proposed  methods  are scalable  to  any  arbitrary  multi-point  configuration  and  dense deployments. Finally, numerical examples are presented to quantify the complexity and the performance advantages of the proposed solutions in terms of achievable sum-rate and reliable connectivity.
\end{itemize}

The paper is an extended version of our previously published conference papers~\cite{Dileep_EUSIPCO2019, Dileep_Globecom2019}. In~\cite{Dileep_EUSIPCO2019} we studied beamformer design for \ac{WSRM} problem leveraging the \ac{SCA} framework while in~\cite{Dileep_Globecom2019} an iterative \ac{KKT} based solution was provided. 
{Compared to the previous work~\cite{Dileep_EUSIPCO2019, Dileep_Globecom2019}, we have included the following additional contributions that provide more complete coverage and analysis.
In this paper, we consider a more practical, spatially correlated and distance-dependent blockage model. Specifically, the presence and/or absence of the blockage on different {RRUs} links are subject to the blockage density, their location, and the resulting distance with the users. In addition, we also provide a closed-form upper bound on the outage performance. 
We further provide detailed complexity analysis for the centralized \ac{SCA} based solution and iterative \ac{KKT} based solution. We also studied and provided extensive simulation on the rate of convergence and its impact on different feasible initialization.
Moreover, the proposed solutions are extended to provide low-complexity two-stage hybrid analog-digital beamforming. These are scalable to any arbitrary multi-point configuration and dense deployments. 
Finally, we have extended the simulation model to take into account user-centric clustering, i.e., we have included the user-RRU association, which results in partially overlapping user-centric clusters. Thus, it leads to inter-cluster and intra-cluster inference conditions. %%%%, and significantly complicates the optimization problem. 
Using this model, we have provided a more extensive set of simulations to illustrate the effectiveness of the proposed methods in terms of achievable sum-rate and reliable \ac{mmWave} connectivity.}
%%%%%All of the aforementioned results have been further improved and extended in this paper.

%\vspace{-10px}
%% ------- Organization and Notations -----------------------------------------
\subsection{Organization and Notations}
\label{subsec:Organization-Notations}
The remainder of this paper is organized as follows. In Section~\ref{sec:model}, we illustrate system, channel, and blockage model as well as, provide the formulation of the problem. Section~\ref{sec:ReliablityVSCoMP} provides a  theoretical analysis of blockage and evaluation of rate and reliability trade-off. In Section~\ref{sec:Precoder-Design}, we describe the robust beamformer designs. The validation of our proposed methods with the numerical results are presented in Section~\ref{sec:Sim-Result}, and finally conclusions are given in Section~\ref{sec:conclusion}. 

\noindent \textit{Notations}: In the following, we represent matrices and vectors with boldface uppercase and lowercase letters, respectively. The transpose, conjugate transpose and inverse operation are represented with the superscript $(\cdot)\tran$, $(\cdot)\herm$ and $(\cdot)^{\scriptsize -1}$ respectively.  $|\mathcal{X}|$ indicates the cardinality of a set $\mathcal{X}$. $\Re\{\cdot\}$ and $|\cdot|$ represent the real part and norm of a complex number, respectively.  $\mathbb{I}_{N}$ indicates the $N\text{x}N$ identity matrix. $\mathbb{C}^{M\text{x} N}$ is a ${M\text{x} N}$ matrix with elements in the complex field. $[\mathbf{a}]_n$ is the $n${th} element of~$\mathbf{a}$. Finally, $\nabla_{\mathbf{x}}y(\mathbf{x})$ denotes gradient of $y(\cdot)$ with respect to variable~$\mathbf{x}$.

%\vspace{-5px}
%% --- System MODEL --------------------------------------------
\section{System Architecture And Problem Formulation}
\label{sec:model}
%
%
%\vspace{-5px}
\subsection{System Model}
\begin{figure}[t]
\centering
\setlength\abovecaptionskip{-0.25\baselineskip}
\includegraphics[trim=0.02cm 0.02cm 0.02cm 0.02cm, clip, width=0.85\linewidth]{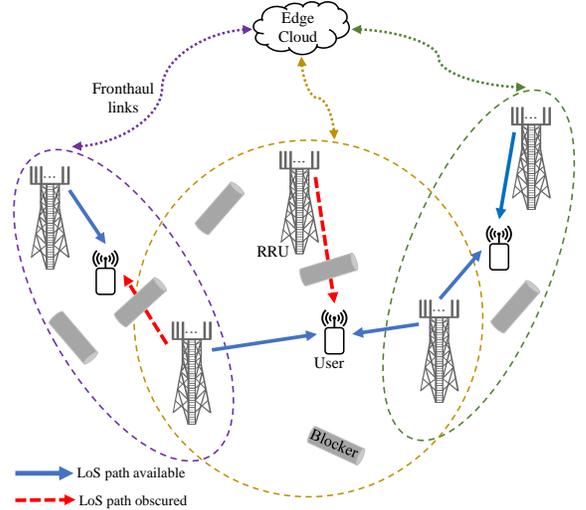} 
\caption{{System model showing transmitters~(RRUs) and receivers~(users) in the presence  of randomly distributed blockers.}} 
\label{fig:System-Model}
\end{figure}

We consider downlink transmission in a \ac{mmWave} based \ac{MU-MISO} communication system, consisting of $K$~single antenna users served by $B$~\acp{RRU}. Each \ac{RRU} is equipped with $N_t$ transmit antennas, and arranged in a \ac{ULA} pattern. The antennas have $0$~dBi gain and $x=\lambda/2$ spacing between any two adjacent elements, where $\lambda$ is the wavelength of carrier frequency.  
We~define $\mathcal{B}=\{ 1, 2,\dots, B \}$ to be the set of all \ac{RRU} indices, $\mathcal{K}=\{ 1, 2,\dots, K \}$ denotes the set of active users, and the serving set of \acp{RRU} for each user~$k$ is represented with $\mathcal{B}_k \subseteq \mathcal{B}$ for all $k\in\mathcal{K}$. %%%Furthermore, we assume, $N_t \geq K$ i.e., all users can be spatially multiplexed.   
We study {\ac{JT}-\ac{CoMP}} transmission, whereby, each active user~$k$ receives a coherently synchronous signal from all the \acp{RRU} in~$\mathcal{B}_k$. Furthermore, the downlink transmissions are performed using the same frequency and time resources. In this paper, if not mentioned otherwise, we assume by default a case where each antenna is connected to a dedicated radio frequency~(RF) chain and baseband circuit that enables fully digital signal processing. In addition, we provide an implementation for two-stage hybrid analog-digital beamforming architecture with coarse-level analog beamforming followed by less-complex digital precoding.
Finally, we assume a cloud (or centralized) radio access network ({C-RAN})~architecture, wherein all RRUs are connected to the edge cloud by high-bandwidth and low-latency fronthaul links, as illustrated in Fig.~\ref{fig:System-Model}. 

It should be noted, in the C-RAN architecture, a common \ac{BBU} performs all the digital signal processing functionalities in a centralized manner, while the RRUs implement limited radio operations~\cite{CRAN-2015}.  Such, fully centralized baseband processing provides more efficient RRU coordination and, thus, enables more effective implementation for \ac{JT}-\ac{CoMP} scenarios~\cite{CRAN-2015}. 
However, in practice, fronthaul link capacity and signaling overhead will limit the maximum number of coordinating \acp{RRU} for each user.
Furthermore, perfect estimation of available \ac{CSI} is assumed at the BBU for the downlink beamformer design and resource allocations, whereby, each RRU receives information for the {active users}, such as control and data signals, using the fronthaul~links. 
%%

%% ----- %%%% ----- %%
The received signal\footnote{The methods can be extended to hybrid analog-digital beamforming architecture where the channel between a RRU-user pair $\{\mathbf{h}_{b,k}\}_{b\in\mathcal{B}, k\in\mathcal{K}} $ can be considered as an effective channel obtained after the analog beamforming stage~\cite{Alkhateeb-LimitedFeedback-2015} (for more details see Section~\ref{subsec:Hybrid}).} of $k$th user, $y_k$ can be expressed as
\begin{equation}
\label{eq:Rx-Signal}
y_k = \sum\limits_{b\in\mathcal{B}_k} \mathbf{h}_{b,k}\herm \mathbf{f}_{b,k} {s_{k}} +   
{ \sum\limits_{u \in \mathcal{K} \setminus k}\sum\limits_{b\in\mathcal{B}_{u}} \mathbf{h}_{b,k}\herm \mathbf{f}_{b,u} {s_{u}} }
+ {w}_k ,
\end{equation}
where $\mathbf{h}_{b,k} \in \mathbb{C}^{N_t \text{x} 1}$ is the channel between a RRU-user pair~$(b,k)$, ${w}_k \sim \mathcal{CN}(0,\sigma_k^2)$ is circularly symmetric additive white Gaussian noise (AWGN) with power spectral density (PSD) of~$\sigma_k^2$ and $s_k$ is normalized and independent data symbol, i.e., $\mathbb{E}\{|s_k|^2\}=1$ and $\mathbb{E}\{s_k s_u^*\}=0$, for all $k,  u \in \mathcal{K}$. In expression~\eqref{eq:Rx-Signal}, $\mathbf{f}_{b,k} \in \mathbb{C}^{N_t \text{x} 1}$ represents the portion of the joint beamformer between a RRU-user pair~$(b,k)$, designed by the centralized BBU assuming perfect estimation of available CSI.
The received \ac{SINR} for each user~$k$ can be expressed~as 
\begin{equation}
\Gamma_k(\mathcal{B}_k) = \frac{\Bigl|\sum\limits_{b\in \mathcal{B}_k}\mathbf{h}_{b,k}\herm \mathbf{f}_{b,k}\Bigr|^2  }{\sigma_k^2 + \sum\limits_{u \in \mathcal{K} \setminus  k} \Bigl| \sum\limits_{b\in \mathcal{B}_{u}}\mathbf{h}_{b,k}\herm \mathbf{f}_{b,u}\Bigr|^2}  .
\label{eq:SINR}
\end{equation}

%
%% ----- %%%% ----- %%
%% ----- %%%% ----- %%
%

%% ----- %%%% ----- %%
%\vspace{-10px}
\subsection{Channel Model}
\label{subsec:channel}
Due to the higher penetration and path-loss, reduced scattering and diffraction at the \ac{mmWave} frequencies compared to sub-$6$~GHz frequency band, the channel can be considered to be spatially sparse~\cite{Rappaport-MillimeterWave-2013, Rappaport-WidebandMillimeter-WavePropagation-2015}, in which \ac{LoS} is the dominant path and mainly contributes to the communication~\cite{MacCartney-RapidFading-2017, AndreevBlockage2018}. Thus, unblocked \ac{LoS} link is highly desirable in order to initiate and maintain a stable \ac{mmWave} communication.   %%%%%%~\cite{Bai-Blockage-2014, Renzo-2015-BinaryChannel}.
{The channel model, in this paper, is based on a sparse geometric model~\cite{Sayeed-SpatialChannel-2002, Clustered-Channel-2016}, which is widely adopted in studies related to mmWave signal processing. The considered channel model is customarily used to model the mmWave radio channels, as it accurately accounts for its high free-space path loss and low-rank nature of mmWave radio signal~\cite{Ayach-SpatiallySparse-2014, Alkhateeb-LimitedFeedback-2015, Yu-Alternating-2016}.} %%% Gao-Energy-Efficient-2016
Specifically, we consider $M_{b,k}$~paths for the channel $\mathbf{h}_{b,k}$ between RRU $b$ and user $k$, and expressed~as
\begin{equation}
\label{eq:channel-def}
\mathbf{h}_{b,k} = \sqrt{\frac{N_t}{M_{b,k}}} \Biggl[  \underbrace{ g_{b,k}^1 \mathbf{a}_T\herm(\phi_{b,k}^1) }_{\mathbf{h}^{\text{LoS}}_{b,k}} + \underbrace{\sum\limits_{m=2}^{M_{b,k}} g_{b,k}^{m} \mathbf{a}_T\herm(\phi_{b,k}^m)}_{\mathbf{h}^{\text{NLoS}}_{b,k}} \Biggr], 
\end{equation}
where $\phi_{b,k}^1$ and $\phi_{b,k}^m$ denote the {\ac{AoA} at $k$th user} for the {LoS} and the $m$-th non-\ac{LoS} (NLoS) path, respectively. Note that the \ac{AoA} for each NLoS path~$m > 1$ is assumed to be uniformly distributed, i.e., $\phi_{b,k}^m \in [-\pi/2, \ \pi/2]$, whereas, the LoS AoA $\phi_{b,k}^1$ is related to the actual position of RRU-user $(b,k)$~pair~\cite{kumar2018reliable}. 
Finally, $g_{b,k}^1=v_{b,k}^{1} d_{b,k}^{-\varrho}$ and $g_{b,k}^{m}=v_{b,k}^{m} d_{b,k}^{-\zeta}$, in which $v_{b,k}^{m} \ \forall m$ is a random complex gain with zero mean and unit variance, $d_{b,k}$ is the RRU-user distance,  $\varrho$ and $\zeta$ denotes the path-loss exponent for the LoS and the NLoS link, respectively. 
It has been shown empirically that $\zeta$~is much higher than the LoS path-loss exponent %$\varrho$
\cite{MacCartney-RapidFading-2017, MacCartney-BaseStationDiversity-2017, MacCartney-BSDiversityJournal-2019}.
The transmit array steering vector of \ac{ULA} for angle $\phi$ is denoted~as $\mathbf{a}_T(\phi) \in \mathbb{C}^{N_t \text{x} 1}$ with
\begin{equation}
\label{eq:array-response}
\big[\mathbf{a}_T(\phi)\big]_n = \frac{1}{\sqrt{N_t}} e^{-\mathsf{j}\frac{2\pi x}{\lambda}(n-1)\sin(\phi)} , \ \ n = 1,2,\ldots,N_t . 
\end{equation}
%

%% ----- %%%% ----- %%
%%\vspace{-10px}
\subsection{Blockage Model}
\label{subsec:blockage}

In the mmWave communication, the quality of the wireless link between a RRU-user pair mainly depends on the characteristics of the \ac{LoS} path~\cite{MacCartney-RapidFading-2017, AndreevBlockage2018}. The {NLoS} paths in the mmWave radio channel are typically $20-30$~dB weaker than the dominant LoS path~\cite{AndreevBlockage2018}, thus high data rates are difficult to achieve in the NLoS only transmission. On the other hand, a major challenge for LoS dominated communication stems from the fact that LoS links may easily be blocked by obstacles. This may result in an intermittent connection, which severely impacts the quality of user-experience. Channel measurements in typical mmWave scenarios have demonstrated that outage on a mmWave link occurs with $20\%-60\%$ probability~\cite{Rappaport-WidebandMillimeter-WavePropagation-2015} and that may lead to over $10-$fold decrease in the achievable rate~\cite{Zhang-Transportlayer-2016}.     
Therefore, unless being addressed properly, the blockage appears as the main bottleneck hindering the full exploitation of the mmWave channel.

%% ----- %%%% ----- %%
%\begin{figure}[t]
%\centering
%\includegraphics[trim=0.2cm 0.1cm 0.2cm 0.1cm, clip, width=\linewidth]{figs/FigPlot3.eps} 
%\caption{The propagation environment between any typical BBU-user pair.} 
%\label{fig:LoSNlOS-Model}
%\end{figure}
%% ----- %%%% ----- %%
%% ----- %%%% ----- %%
\begin{figure}[t]
\centering
\setlength\abovecaptionskip{-0.25\baselineskip}
\includegraphics[trim=0.2cm 0.2cm 0.2cm 0.1cm, clip, width=0.80\linewidth]{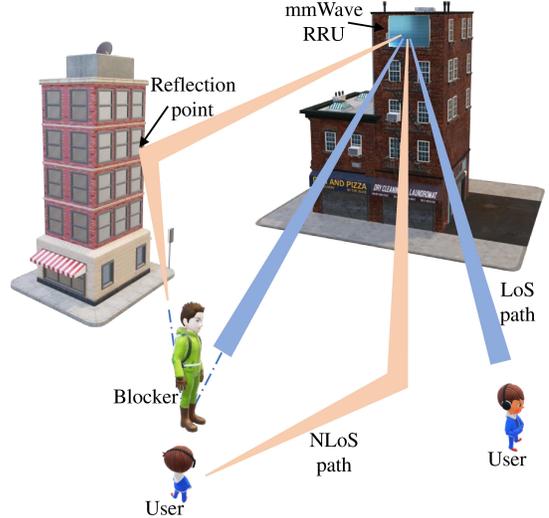} 
\caption{The downlink propagation environment between a  RRU-user~pair.} 
\label{fig:LoSNlOS-Model}
\end{figure}
%% ----- %%%% ----- %%

%
{In order to characterize the aforementioned uncertainties of the mmWave radio channel, we consider probabilistic blockage model, where the link specific blockage depends on the blockage density and the distance between each RRU-user pair%%%%, where each \revise{path between RRU-user pair is assumed to have an independent blockage probability}. This is a reasonably accurate assumption especially when the blockers are not overly~large and close to the users
~\cite{Bai-Blockage-2014, Renzo-2015-BinaryChannel}.  
%%%%%Further, due to high penetration loss and weak diffraction of mmWave signal, the blockage sources can be assumed to be impenetrable~\cite{Maamari-CoverageinmmWave-2016}, namely, communications can be established only in the presence of the unblocked paths.       
%
For simplicity, we assume independent blockage events per link\footnote{{This is an accurate assumption especially when the blockers are not overly large and not too close to the users~\cite{Bai-Blockage-2014, Renzo-2015-BinaryChannel}.}} only for the dominant LoS path while all the NLoS links are assumed to be unobstructed\footnote{{The methodology can be extended to more elaborate blocking models, e.g., by considering the spatio-temporal correlation between the LoS/NLoS states and also across the coordinating RRUs~\cite{Hriba-CorrelatedBlocking-2019}. In fact, this is an interesting topic for future extension.}}. More specifically, the channel between any typical RRU-user pair can be either fully-available or in NLoS state.} The~NLoS state occurs when the dominant LoS link is blocked by any obstacle. It should be noted that even a mobile human blocker may cause $20-30$~dB attenuation and can obstruct the LoS path for hundreds of milliseconds~\cite{MacCartney-RapidFading-2017, AndreevBlockage2018}. This can be equivalently modeled as~$\{\mathbf{h}^{\text{LoS}}_{b,k}=\mathbf{0}\}_{b\in\mathcal{B},  k\in\mathcal{K}}$ for the blocked LoS component. The fully-available state is defined~in~\eqref{eq:channel-def}.

{Since the blockers are completely random, their position and orientation 
%%%%%%may change multiple times within the channel coherence interval. Further, the blockage events 
cannot be known a priori in a dynamic mobile environment.}  
Similar to~\cite{Bai-Blockage-2014, Renzo-2015-BinaryChannel}, the blockage between RRU-user pair $(b,k)$ is defined using a Bernoulli random variable with the parameter $\eta$ and expressed~as
\begin{equation}
\label{eq:blockage_model}
\mathbf{h}^{\text{LoS}}_{b,k} = \left\{
    \begin{array}{ll}
        g_{b,k}^1 \mathbf{a}_T\herm(\phi_{b,k}^1) & \text{with probability }  e^{-\eta d_{b,k}} \\
        \mathbf{0} & \text{with probability }  1 - e^{-\eta d_{b,k}}
    \end{array}
\right.
\end{equation}
where $\eta$ depends on the density and the average size of the obstacles blocking the dominant LoS path. In addition, the probability of a LoS link decreases exponentially with the increase in distance between RRU-user~\cite{Bai-Blockage-2014, Renzo-2015-BinaryChannel}. 
{{From the physical standpoint, $\eta$ can be interpreted as the LoS likelihood for a given propagation environment and distance~\cite{Bai-Blockage-2014}.
%%%%%, i.e., the smaller the $\eta$ value the sparser the propagation environment. Consequently, there is higher probability of a LoS link at the given distance and vice-versa. 
%
For example, the smaller the $\eta$ value, the sparser the propagation environment, and consequently, the higher probability of a LoS link at the given distance and vice-versa. 
In our study, we will make use of parameter~$\eta$ to analyze the effect of different blockage densities on the system performance.}  
On the other hand, for a fixed blockage density $\eta$ (which is common for all RRUs in the system), the LoS blockage is only subject to the link distance $d_{b,k}$ between any typical RRU-user pair for all $ b\in\mathcal{B}$ and $k\in\mathcal{K}$. Thus, the presence and/or absence of the dominant LoS path is subject to the blockage density and the distance between the RRUs and the users.}
%
%%%%%%%%%%%%%%An~additional inaccessible state may occur if the corresponding path-loss becomes sufficiently high to establish a communication link~\cite{Renzo-2015-BinaryChannel}.       

{
We assume a standard \ac{TDD}-based \ac{CSI} acquisition from reciprocal uplink followed by the downlink data transmission phase. More specifically, BBU designs the transmit beamformer based on the available \ac{CSI} acquired during the (uplink) estimation phase~\eqref{eq:channel-def}. 
Thus, a system can be in the outage, if the dominant LoS link was available during the channel estimation, and is not anymore available during the data  transmission phase due to random blockage (e.g., due to channel aging of blockage effect). 
Similarly, a LoS link can also be in the blockage state during the channel estimation phase. However, these links will not be included for the downlink data transmission. As a consequence, the actual achievable rate at the receiver would be larger than the assigned rate to the users.
Thus, from the reliability perspective, we have to consider the case when the channel is available during the estimation phase but blocked during the data transmission phase, which is not known at the BBU for downlink beamformer~design.}

%% ----- %%%% ----- %%
%%\vspace{-10px} 
\subsection{Problem Formulation}
\label{subsec:problem}
The major goal of this work is to develop a robust and resilient downlink transmission strategy that can retain stable connectivity under the uncertainties of mmWave radio channel and random  blockages. 
To this end, we need to compute the optimal joint transmit beamformer $\mathbf{F}~=~[\mathbf{f}_{1,1}, \mathbf{f}_{1,2}, \dots, \mathbf{f}_{B,K}]$, while exploiting the multi-antenna spatial diversity and \ac{CoMP} connectivity for improved system-level reliability.     
For the \ac{WSRM}\footnote{We assume Gaussian signalling as the upper bound for the rate expressions.  Further, each user decodes its intended signals by treating all other interfering signals as noise in~\eqref{eq:SINR}.
} 
  objective considered in this paper\footnote{The formulation can easily be applied to other objective functions, e.g., minimum user-rate maximization~\cite{Dileep_EUSIPCO2019} or weighted queue minimization~\cite{Ganesh-TrafficAware-2016}.}, the beamformer design can be formulated~as 
%%%%%
%
\begin{subequations}
\label{eq:P1-objective}
\begin{align}
	\displaystyle 
	    \underset{\mathbf{F}, \tilde{\gamma}_k}{\text{maximize}} \quad &\sum\limits_{k\in \mathcal{K}} \delta_k 
		\log \big(1+\tilde{\gamma}_k\big) \\
		\text{subject to} \quad
	&\displaystyle {\Gamma}_{k}({\mathcal{B}}_k\comb) \geq \tilde{\gamma}_k \ \
	  \forall k, \ \forall {\mathcal{B}}_k\comb \subseteq \mathcal{B}_k, \ |{\mathcal{B}}_k\comb|\geq L_k
	 \label{eq:P1-C1} ,
 \\
	&\displaystyle \sum\limits_{k \in \mathcal{K}}\|\mathbf{f}_{b,k}\|^2 \leq P_b  \ \ \forall b \label{eq:P1-C2} , 
\end{align}
\end{subequations}
{where $\delta_k \geq 0 \ \forall k$ denotes the user-specific priority weights corresponding to the achievable rate, and they are fixed before data transmission (e.g., by BBU).} \revise{The function ${\Gamma}_{k}(\cdot)$ is defined in expression~\eqref{eq:SINR}.  For each user $k$, the constraint~\eqref{eq:P1-C1} is the pessimistic estimate of \ac{SINR} computed over all possible subset combinations of potentially available RRUs, each of size {$|{\mathcal{B}}_k\comb|\geq L_k$}, from its serving set~$\mathcal{B}_k\comb (\subseteq \mathcal{B}_k)$, e.g., by excluding the blocked \acp{RRU}, {as will become clear in following}.} The total transmit power for $b$th RRU  is bounded by~$P_b$, as in~\eqref{eq:P1-C2}. 

In practice, the adverse channel condition and signaling overhead limits the maximum number of cooperating \acp{RRU} for each user (i.e., $\mathcal{B}_k \ \forall k$)~\cite{CRAN-2015}. Thus, the subset combinations $\mathcal{B}_k\comb (\subseteq \mathcal{B}_k)$ are fairly small for modestly sized systems. %%%%Hence, the problem~\eqref{eq:P1-objective}, in general, does not require combinatorial optimization.
The resulting problem~\eqref{eq:P1-objective} is intractable due to non-convex and coupled \ac{SINR} constraints~\eqref{eq:P1-C1}. 
%
%%%In addition, the number of \ac{SINR} constraint combinations can be large depending on $|{\mathcal{B}}_k\comb|$ and $L_k$. 
%
To this end, in Section~\ref{sec:Precoder-Design}, we provide practical and computationally efficient iterative algorithms by exploiting convex approximation techniques.

%% --- ReliablityVSCoMP --------------------------
%%\vspace{-10px} 
\section{Analysis of Rate and Reliability Trade-off}
\label{sec:ReliablityVSCoMP}
In this paper, we assume randomly distributed blockers. Thus, the position of each blocker and/or blockage event is completely unknown. {Therefore, to improve system reliability and avoid outage under the uncertainties of mmWave radio channel, we preemptively underestimate the achievable SINR assuming that a portion of available CoMP links would be blocked during the data transmission phase. This is specifically required in the mmWave, because of dynamic blockages which are not possible to track during the channel estimation phase. 
Let BBU assumes that each user~$k$ have at least $L_k $ available links (i.e., unblocked RRUs). Then, the BBU proactively models the SINR over all possible subset combinations, by excluding the potentially blocked RRUs, and allocate the rate to users such that transmission reliability is improved (i.e., minimize the outage due to blockages that appear during the data transmission phase).}

%%%and uses the worst case estimate of~$\tilde{\gamma}_k$, in the objective, for the beamformer design.
%
For example, referring to expression~\eqref{eq:P1-C1} and Fig.~\ref{fig:System-Model}, 
let $\mathcal{B}_k$ be the set of \acp{RRU} that are used to serve $k$th user with RRU indices $\mathcal{B}_k=\{1,2,3,4\}$. 
Then, with the assumption of at least $L_k=3$ available links, 
%%%%all the possible subset combination~${\overline{\mathcal{B}}}_{k}$ of unblocked \acp{RRU}  available to $k$th user include the following indices:
the serving set of unblocked \acp{RRU} available to $k$th user can be any one of the following combinations:
%%%%%%%%%%%%%%let the $k$th user have  $|\mathcal{B}_k|=4$ and $L_k=3$, then all the possible subset combination~${\overline{\mathcal{B}}}_{k}$ of available RRUs include the following indices
%
\begin{align*}
%\label{eq:BS_Combination}
\overline{\mathcal{B}}_{k} = \big\{\{1, 2, 3\}, \{1, 2, 4\}, \{1, 3, 4\}, \{2, 3, 4\}, \{1, 2, 3, 4\}\big\}.
\end{align*}
Equivalently, the subset combinations of all blocked RRUs $\overline{\mathcal{D}}_k$ for the $k$th user can be expressed~as
\begin{equation*}
    \overline{\mathcal{D}}_k  = \big\{ \{4\}, \{3\}, \{2\}, \{1\}, \{\} \big\}.  %%%\triangleq \mathcal{B}_k \backslash \overline{\mathcal{B}}_k
\end{equation*}

Let $C(L_k)$ denotes the cardinality of set $ \overline{\mathcal{B}}_k$ and defined as  $C(L_k)=\sum_{l=L_k}^{|{\mathcal{B}}_k|} \frac{|{\mathcal{B}}_k|!}{l! (|{\mathcal{B}}_k| - l)!}$. 
%
%%%%%%In practice, the adverse mmWave channel condition and signaling overhead limits the maximum number of cooperating \acp{RRU} for each user (i.e., $\mathcal{B}_k \ \forall k$)~\cite{CRAN-2015}. Thus, $C(L_k)$ is fairly small for modestly sized systems. Hence, the problem~\eqref{eq:P1-objective}, in general, does not require combinatorial optimization. 
%
Recall that, the set of coordinating \acp{RRU} for each user (i.e., $|\mathcal{B}_k| \ \forall k$) are limited. Thus, $C(L_k)$  is fairly small and solving~\eqref{eq:P1-objective}, in general, does not require combinatorial optimization.
Let $\mathcal{D}_k\comb\in\overline{\mathcal{D}}_k$ denotes $c$-th subset combination of the potentially blocked RRUs, and $\mathcal{B}_k\comb \in \overline{\mathcal{B}}_k$ represents $c$-th subset combination of the  available RRUs for $k$th user, where $c=1,2,\dots,C(L_k)$. 
Then, the SINR of each user $k$ for $c$-th subset combination (i.e., $\mathcal{B}_k\comb \in \overline{\mathcal{B}}_{k}$) is obtained by excluding the blocked RRUs in~\eqref{eq:SINR} and expressed~as
\begin{equation}
\Gamma_k(\mathcal{B}_k\comb) = \frac{\Bigl|\sum\limits_{b\in \mathcal{B}_k\comb}\mathbf{h}_{b,k}\herm \mathbf{f}_{b,k}\Bigr|^2  }{\sigma_k^2 + \sum\limits_{u \in \mathcal{K} \setminus  k} \Bigl| \sum\limits_{b\in \mathcal{B}_{u} \setminus \mathcal{D}_k\comb}\mathbf{h}_{b,k}\herm \mathbf{f}_{b,u}\Bigr|^2}  .
\label{eq:SINR-comb}
\end{equation}
%
%%%%where $\mathcal{D}_k\comb = \mathcal{B}_k \backslash \mathcal{B}_k\comb$ denotes $c$-th subset of potentially blocked \acp{RRU} which are excluded from the interfering links to $k$th~user. 
%
{where $\mathcal{B}_k\comb = \mathcal{B}_k \backslash \mathcal{D}_k\comb$ and $\mathcal{D}_k\comb\in\overline{\mathcal{D}}_k$ denotes $c$-th subset of potentially blocked \acp{RRU}%%%which are also excluded from the interfering links to $k$th~user
.}
Consequently, after solving the problem~\eqref{eq:P1-objective}, the pessimistic estimate of the achievable SINR for $k$th user is equivalent to $\tilde{\gamma}_k\triangleq\underset{c=1, \ldots, C(L_k)}{\mathrm{min}}\big( \Gamma_k({\mathcal{B}}_{k}\comb) \big)$ for all $k\in\mathcal{K}$. 
Still, the data symbols $s_k$ are transmitted to each user~$k$ from all $\mathcal{B}_k ( \supseteq {\mathcal{B}}_k\comb)$ RRUs. 
Therefore, {reliable connectivity for each user~$k$ can be guaranteed, even if, $|\mathcal{B}_k| - L_k$ dominant links are not available during the transmission phase. Contrarily, if~more than $L_k$ links were available, the actual achievable rate would be larger than the assigned rate to the~users}.

%%%%%%%%%%%%%%%%%%%%%%%%%%%%%%%%%%%%%%%%%%%%%%%
%
%
\begin{comment}
As an example, let $q_k$ denotes the average blocking probability of $k$th user and defined as $q_k = \frac{1}{|\mathcal{B}_k|} \sum_{b\in\mathcal{B}_k} 1 - e^{-\eta d_{b,k}} $ for all $k \in \mathcal{K}$. Next, by using the binomial distribution, we can approximately model the success probability $\widetilde{q}_k$ for $k$th user~as
%
\begin{equation}
    \label{eq:Pout_user}
    \widetilde{q}_k \big(|\mathcal{B}_k| ,q_{k}\big) = \sum\limits_{t=0}^{|\mathcal{B}_k|-L_k} \binom{|\mathcal{B}_k|}{t} (1-q_{k})^{|\mathcal{B}_k|-t} q_{k}^t \ \ \forall k.
\end{equation}
%
%
%%%%%
%
%
As an example, let the average blocking $q_{k}~=~\frac{1}{|\mathcal{B}_k|}\sum_{b\in\mathcal{B}_k} 1 - e^{-\eta d_{b,k}}$. Then, by using the binomial distribution, we can approximately model the success probability $\tilde{q}_k$ for user~$k$~as
%
\begin{equation}
    \label{eq:Pout_user}
    \tilde{q}_k \big(|\mathcal{B}_k| ,q_{k}\big) = \sum_{t=0}^{|\mathcal{B}_k|-L_k} \binom{|\mathcal{B}_k|}{t} (1-q_{k})^{|\mathcal{B}_k|-t} q_{k}^t  \ \forall k,
\end{equation}
%
where $q_{b,k} \in [0 ,1]$ is the probability of a channel is in the blockage state. 
%
\end{comment}
%
%
%
%%%%%%%%%%%%%%%%%%%%%%%%%%%%%%%%%%%%%%%%%%%%%%%

As an example, let $q_{b,k}$ represent the blockage probability between RRU-user pair $(b,k)$, defined as $q_{b,k}=1-e^{-\eta d_{b,k}}$ (see \eqref{eq:blockage_model}). Next, for a given parameter $L_k$, we can approximately model the success probability\footnote{If we assume equal blockage probability i.e.,  $q_{b,k}=q_k \ \forall b\in\mathcal{B}_k$, then success probability of $k$th user $p_k$ becomes a binomial distribution~\cite{Dileep_EUSIPCO2019} and can be expressed as $p_k=\sum_{t=0}^{|\mathcal{B}_k|-L_k} \binom{|\mathcal{B}_k|}{t} (1-q_{k})^{|\mathcal{B}_k|-t} q_{k}^t\ \forall k\in\mathcal{K}$.} $p_k$ of $k$th user as
\begin{equation}
    \label{eq:SuccessProb}
    p_k = \sum\limits_{c=1}^{C(L_k)} \bigg( \prod_{b\in\mathcal{B}_k\comb} (1-q_{b,k}) \times \prod_{b\in\mathcal{D}_k\comb} q_{b,k} \bigg) \ \ \forall k.
\end{equation}
%
%
%%%%An upper bound of expression~\eqref{eq:SuccessProb} is provided in Appendix~\ref{app:success}. 
In Appendix~\ref{app:success}, we generalize~\eqref{eq:SuccessProb} by integration over random users location. 

Since all users are independent, therefore, the system is in outage if any of the $K$ users is in outage. It should be noted, this is a worst-case assumption to enforce strict system-level~reliability. However, in practice, users with the unblocked links can still decode their received signal. Thus, with the worst-case assumption, system outage is defined~as  
\begin{equation}
\label{eq:Pout_Formula}
	\widetilde{P}_{\mathrm{out}} = 1 - \prod_{k=1}^K p_k .
\end{equation}
{The closed-form expression~\eqref{eq:Pout_Formula} models the case when the channel between a RRU-user pair is either fully-available or completely-blocked (e.g., both LoS and NLoS paths in~\eqref{eq:channel-def}), which corresponds to the local-blockage (or self-blockage) at the user. However, we consider blocking of the dominant \ac{LoS} link while keeping all NLoS components unobstructed~(see Section~\ref{subsec:blockage}). Thus, expression~\eqref{eq:Pout_Formula} provides an approximation on the outage performance, as shown in Section~\ref{sec:Sim-Result}.}

Intuitively, we can observe the impact of constraint~\eqref{eq:P1-C1} on the system reliability and achievable rate using~\eqref{eq:SINR-comb} and \eqref{eq:Pout_Formula}. {For example, by using the smaller subset size (i.e., parameter $L_k \ \forall k$), we can improve the system reliability assuming that a significant portion of all the available CoMP links (i.e., $|\mathcal{B}_k|-L_k \ \forall k$) are potentially blocked. However, it leads to a lower SINR estimate and, hence, a lower rate to each user. Conversely, a less pessimistic assumption  on subset size can provide the higher instantaneous SINR and user-specific rates, but it is more susceptible to the outage, thus resulting in less stable connectivity for each user.} Clearly, there is a trade-off between achievable rate and reliable~connectivity\footnote{The user-specific subset size $L_k$ {and the serving set size $|\mathcal{B}_k| \ \forall k\in\mathcal{K}$ are} design parameters that can be tuned based on statistical information, e.g., users location and blockage density, to achieve a desired rate-reliability trade-off (for more details see Section~\ref{subsec:sum-rate-Results}).}. %%%%%Our goal is to quantify this trade-off.
%
%%%%%%%%Not that, user-specific subset size (i.e., $L_k \ \forall k$) is a design parameter based on statistical information, e.g., users location and blockage density, and tuned to achieve desired rate-reliability trade-off. However, dynamic optimization of subset size is beyond the scope of this paper and left for future work. 
%
%%%%% \footnote{The success probability can be upper bounded by using the by using the binomial distribution, and expressed as $$ {p}_{k} = \big( |\mathcal{B}_k|- \Psi\big)\binom{|\mathcal{B}_k|}{\Psi} \frac{1}{|\mathcal{B}_k|} \int_{b\in\mathcal{B}_k}^{} \int_{0}^{1-\overline{q}_{b,k}} t^{|\mathcal{B}_k|-\Psi-1}(1-t)^{\Psi} dt  ,$$ where $\Psi = |\mathcal{B}_k|-L_k$. The mean blocking probability $\overline{q}_{b,k}$ between a BS-user pair $(b,k)$ is defined as $$ \overline{q}_{b,k} = \frac{1}{2x_k} \frac{1}{2y_k} \int_{-x_k}^{x_k} \int_{-y_k}^{y_k} 1 - e^{-\eta d_{b,k}} dx dy $$ }

%% --- Precoder Desing --------------------------
%%\vspace{-10pt}
%\section{Proposed Beamformer Designs For Reliable Communication}
\section{Proposed Beamformer Designs}
\label{sec:Precoder-Design}
In this section, we elaborate on solving problem~\eqref{eq:P1-objective}, which is intractable as-is, mainly due to non-convex \ac{SINR} in~\eqref{eq:P1-C1}.
{Several approaches have been outlined in the existing literature to handle the \ac{SINR} non-convexity, e.g., semidefinite relaxation (SDR) technique \cite{Karipidis-SDR-2008}, fractional programming (FP) based quadratic transform~\cite{Yu-FractionalProgram-2018, Yu-FractionalProgram-2020} and linear Taylor series approximation using the \ac{SCA} framework~\cite{marks1978general, beck2010sequential}.}
In this paper, we employ \ac{SCA} framework~\cite{marks1978general, beck2010sequential}, wherein all non-convex constraints are approximated with the sequence of convex approximations. The underlying approximated subproblem is then iteratively solved until convergence.  
{The \ac{SCA} based solutions have been widely used in many practical applications, e.g., in satellite system~\cite{Wang_Satellite_2019}, wire-line DSL network~\cite{Papandriopoulos_2009}, small-cell heterogeneous network~\cite{Ngo_2014}, energy efficiency~\cite{Tervo-EE-2017, Tervo-EE-2018}, spectrum sharing~\cite{Wang-2012} and multi-antenna interference coordination~\cite{Wang-CoMP-2016, Ganesh-TrafficAware-2016, Kaleva-DecentralizeJP-2018}.
In view of the prior works~\cite{Wang_Satellite_2019, Papandriopoulos_2009, Ngo_2014, Tervo-EE-2017, Tervo-EE-2018, Wang-2012, Wang-CoMP-2016, Ganesh-TrafficAware-2016, Kaleva-DecentralizeJP-2018}, there lacks a systemic approach for the design of downlink beamformer in the \ac{JT}-\ac{CoMP} scenario, while accounting the uncertainties of mmWave radio channel and random blockers, thus motivating the current~work.}

%% ----- %%%% ----- %%
\vspace{-10pt}
\subsection{Solution via Successive Convex Approximation}
\label{subsec:SCA}

The non-convex SINR constraint~\eqref{eq:P1-C1} can be handled by \ac{SCA} framework, as shown in~\cite{Wang-CoMP-2016, Ganesh-TrafficAware-2016, Tervo-EE-2017, Kaleva-DecentralizeJP-2018}, where the authors provided SINR approximation method for \ac{CB}~\cite{Wang-CoMP-2016, Ganesh-TrafficAware-2016, Tervo-EE-2017} and \ac{JT}~\cite{Kaleva-DecentralizeJP-2018} assuming global-CSI and no blockage. 
We extend these approaches to take into consideration coherent multi-point transmission and provide a novel grouping of a multitude of potentially coupled and non-convex \ac{SINR} conditions that raise from the link blockage subsets. 
%
%%%%%%%%The individual SINR constraint for each ${\mathcal{B}}_k\comb \subseteq \mathcal{B}_k$ can be solved as in~\cite{Ganesh-TrafficAware-2016}. 
%%%%%%%%The exact details are omitted due to lack of space (we refer the reader to~\cite{Ganesh-TrafficAware-2016} for the details).
{In~the~following, the main steps are briefly described (we refer the reader to~\cite{Ganesh-TrafficAware-2016} for the details).}
To begin with, by using the expression of $\Gamma_k(\mathcal{B}_k\comb)$~(see \eqref{eq:SINR-comb}) and adding one on both sides, we rewrite~\eqref{eq:P1-C1}~as
%
%%%%%%%%%%%%%%%%%%%%%%%%%%  %%%%%%%%%%%%%%%%%%%%%%%%%%
%
%
%
\begin{subequations}
\label{eq:SINR+1}
\begin{align}
& \ 1+\tilde{\gamma}_k  \leq  \frac{ \sigma_k^2 + \sum\limits_{j\in\mathcal{K}} \Bigl|\sum\limits_{b\in {\mathcal{B}}_j \backslash \mathcal{D}_k\comb}\mathbf{h}_{b,k}\herm \mathbf{f}_{b,j} \Bigr|^2  }{\sigma_k^2 + \sum\limits_{u \in \mathcal{K} \setminus  k} \Bigl| \sum\limits_{b\in {\mathcal{B}}_u \setminus \mathcal{D}_k\comb}\mathbf{h}_{b,k}\herm \mathbf{f}_{b,u}\Bigr|^2},
\label{eq:SINR-numa-1} \\
& \ \ =   \frac{ \sigma_k^2 + \sum\limits_{j\in\mathcal{K}} \Bigl|\sum\limits_{b\in {\mathcal{B}}} \big( \mathbbm{1}_{\mathcal{G}_k\comb}(b) \mathbf{h}_{b,k}\herm \big) \big( \mathbbm{1}_{\mathcal{B}_j}(b) \mathbf{f}_{b,j} \big) \Bigr|^2  }{\sigma_k^2 + \sum\limits_{u \in \mathcal{K} \setminus  k} \Bigl| \sum\limits_{b\in {\mathcal{B}}} \big( \mathbbm{1}_{\mathcal{G}_k\comb}(b) \mathbf{h}_{b,k}\herm \big) \big( \mathbbm{1}_{\mathcal{B}_u}(b) \mathbf{f}_{b,u} \big) \Bigr|^2},
\label{eq:SINR-numa-Indicator}
\end{align}
\end{subequations}
where ${\mathcal{G}_k\comb} = {\mathcal{B} \backslash \mathcal{D}_k\comb}$ for all $c=1,2,\ldots, C(L_k)$ and $k\in\mathcal{K}$. The indicator function $\mathbbm{1}_{\mathcal{G}_k\comb}(b)$ and $\mathbbm{1}_{\mathcal{B}_j}(b)$ are defined~as
\begin{eqnarray*}
%\begin{equation*}
\mathbbm{1}_{\mathcal{G}_k\comb}(b) & = & \left\{
    \begin{array}{ll}
        {1} & \text{ if and only if } \ \  b \in \mathcal{B}\setminus\mathcal{D}_k\comb \\
        {0} & \text{ otherwise }
    \end{array}
\right.
%\end{equation*}
%
\\
%%%and
%
%\begin{equation*}
\mathbbm{1}_{\mathcal{B}_j }(b) & = & \left\{
    \begin{array}{ll}
        {1} & \text{ if and only if } \ \   b \in \mathcal{B}_j \\
        {0} & \text{ otherwise }
    \end{array}
\right.
%\end{equation*}
\end{eqnarray*}
Furthermore, the expression~\eqref{eq:SINR-numa-Indicator} can be compactly expressed using the vector notations.  
Let $\overline{\mathbf{f}}_{j}~\in~\mathbb{C}^{ | {\mathcal{B}} | N_t \times 1}$ be the stacked downlink beamformer defined as
\begin{align*}
    \overline{\mathbf{f}}_{j}  & \triangleq  \big[\mathbbm{1}_{\mathcal{B}_j }(1)\mathbf{f}_{1,j}\tran, \dots, \mathbbm{1}_{\mathcal{B}_j }(b)\mathbf{f}_{b,j}\tran, \dots, \mathbbm{1}_{  \mathcal{B}_j}(B)\mathbf{f}_{B,j}\tran\big]\tran,
\end{align*}
and $\overline{\mathbf{h}}_{k}\comb~\in~\mathbb{C}^{ | {\mathcal{B}} | N_t \times 1}$ be the stacked channel vector defined~as 
\begin{align*}
    \overline{\mathbf{h}}_{k}\comb & \triangleq \big[\mathbbm{1}_{\mathcal{G}_k\comb}(1)\mathbf{h}_{1,k}\tran, \dots, \mathbbm{1}_{\mathcal{G}_k\comb}(b)\mathbf{h}_{b,k}\tran, \dots, \mathbbm{1}_{\mathcal{G}_k\comb}(B)\mathbf{h}_{B,k}\tran \big]\tran.
\end{align*}
For the brevity of mathematical representation, we also define $\mathbf{h}_{b,k}\comb = \mathbbm{1}_{\mathcal{G}_k\comb}(b)\mathbf{h}_{b,k}$ for all $b \in \mathcal{B}$.
Thus, by using the vector notations $\overline{\mathbf{f}}_{j}$ and $\overline{\mathbf{h}}_{k}\comb$,  expression~\eqref{eq:SINR+1} can be rewritten~as
\begin{equation}
    \label{eq:Compact-SINR+1}
    1+\tilde{\gamma}_k \leq \frac{ \sigma_k^2 + \sum\limits_{j\in\mathcal{K}} \bigl| \overline{\mathbf{h}}_{k}\cherm \overline{\mathbf{f}}_{j} \bigr|^2  }{\sigma_k^2 + \sum\limits_{u \in \mathcal{K} \setminus  k} \bigl| \overline{\mathbf{h}}_{k}\cherm \overline{\mathbf{f}}_{u}\bigr|^2}.
\end{equation}
Note that we have added one on both sides of constraint~\eqref{eq:P1-C1} to get~\eqref{eq:Compact-SINR+1}. This improves the numerical stability of the algorithm as will become clear in the following.

For more compact representation, we now introduce function $I_k({\mathcal{B}}_k\comb)$ and $\mathcal{H}_k({\mathcal{B}}_k\comb)$, defined as
\begin{subequations}
\label{eq:Function-I-G}
\begin{align}
    I_k({\mathcal{B}}_k\comb) & =  \sigma_k^2 + \sum_{u \in \mathcal{K} \setminus  k} \bigl| \overline{\mathbf{h}}_{k}\cherm \overline{\mathbf{f}}_{u}\bigr|^2, 
\label{eq:SINR-dena}\\
    \mathcal{H}_k({\mathcal{B}}_k\comb) & = \frac{\sigma_k^2 + \sum_{j\in\mathcal{K}} \bigl| \overline{\mathbf{h}}_{k}\cherm \overline{\mathbf{f}}_{j} \bigr|^2}{1+\tilde{\gamma}_k} , 
    %
    %%\mathcal{H}_k({\mathcal{B}}_k\comb) & = {\sigma_k^2 + \sum\limits_{j\in\mathcal{K}} \bigl| \overline{\mathbf{h}}_{k}\cherm \overline{\mathbf{f}}_{j} \bigr|^2} \big/ {1+\tilde{\gamma}_k} , 
\label{eq:SINR-Stacked} 
\end{align}
\end{subequations}
Then expression~\eqref{eq:Compact-SINR+1} can be written as
\begin{equation}
\label{eq:SINR-difference}
     I_k({\mathcal{B}}_k\comb) - \mathcal{H}_k({\mathcal{B}}_k\comb) \leq 0,
\end{equation}
for all   $c = 1,2,\dots C(L_k)$ and $k \in \mathcal{K}$.

Note that function $\mathcal{H}_k({\mathcal{B}}_k\comb)$ is a \emph{quadratic-over-linear}, which is a convex function~\cite[Ch. 3]{boyd2004convex}. 
Hence, the reformulated SINR constraint~\eqref{eq:SINR-difference} is still non-convex (i.e., difference of convex functions). 
{Therefore, we resort to the \ac{SCA} framework~\cite{marks1978general, beck2010sequential}, wherein all non-convex constraints are approximated with a sequence of convex subsets and iteratively solved until convergence of the objective~\cite{Wang-CoMP-2016, Ganesh-TrafficAware-2016, Tervo-EE-2017, Kaleva-DecentralizeJP-2018, Tervo-EE-2018}. Note that the \ac{SCA} framework based on iterative relaxation of non-convex \ac{SINR} constraints can be shown to converge to a stationary point~\cite[Appendix A]{Ganesh-TrafficAware-2016}.} 
Thus, the best convex approximation of reformulated SINR constraint~\eqref{eq:SINR-difference} is obtained by replacing $\mathcal{H}_k({\mathcal{B}}_k\comb)$ with its first-order approximation.
%
%%%%The RHS of~\eqref{eq:SINR-Stacked} can be bounded by the linear first-order Taylor approximation~as
The linear first-order Taylor approximation of $\mathcal{H}_k({\mathcal{B}}_k\comb)$ can be expressed~as
\begin{align}
\label{eq:Linear-Approximation}
& \mathcal{F}\big(c, \overline{\mathbf{f}}_{k}, \tilde{\gamma}_k; \overline{\mathbf{f}}_{k}\iLoop, \tilde{\gamma}_k\iLoop \big) \triangleq  
2 \sum\limits_{j\in\mathcal{K}} \Re  \bigg\{ \frac{\overline{\mathbf{f}}_{j}\iherm \overline{\mathbf{h}}_{k}\comb \overline{\mathbf{h}}_{k}\cherm }{1 + \tilde{\gamma}_k\iLoop}
\big(\overline{\mathbf{f}}_{j} -  \overline{\mathbf{f}}_{j}\iLoop \big) \bigg\} \ \nonumber  \\ & \quad %%\qquad \qquad \qquad \qquad
+ \frac{ \sigma_k^2 + \sum\limits_{j\in \mathcal{K}} \bigl|\overline{\mathbf{h}}_{k}\cherm \overline{\mathbf{f}}_{j}\iLoop\bigr|^2}{1 + \tilde{\gamma}_k\iLoop} 
\bigg( 1- \frac{\tilde{\gamma}_k -\tilde{\gamma}_k\iLoop}{1 + \tilde{\gamma}_k\iLoop} \bigg) 
%
%%\\ & \qquad \qquad \leq 
\leq \mathcal{H}_k({\mathcal{B}}_k\comb) ,
%
%%%\frac{ \sigma_k^2 + \sum\limits_{j\in \mathcal{K}} \bigl| \overline{\mathbf{h}}_{k}\cherm \overline{\mathbf{f}}_{j} \bigr| ^2}{1 + \tilde{\gamma}_k} \nonumber ,
\end{align}
%
%%where \eqref{eq:Linear-Approximation} is the under-estimator for the RHS of~\eqref{eq:SINR-Stacked}, 
with equality only at the approximation point $\{ \overline{\mathbf{f}}_{k}\iLoop, \tilde{\gamma}_k\iLoop\}$. 
After replacing~\eqref{eq:SINR-Stacked} with its linear approximation~\eqref{eq:Linear-Approximation} and plugging it into constraint~\eqref{eq:P1-C1}, an approximated subproblem for $i${th} SCA iteration is expressed in convex form along with the corresponding dual-variables~as 
\begin{subequations}
\label{eq:P1-KKT-objective2}
\begin{align}
	\displaystyle 
	    \underset{\mathbf{F}, \tilde{\gamma}_k}{\text{maximize}} \quad &\sum\limits_{k\in \mathcal{K}} \delta_k 
		\log \big(1+\tilde{\gamma}_k\big) \\
		\text{subject to} \quad 
	\begin{split}
	&\displaystyle a_{k,c}: \ \  \mathcal{F}\big(c,\overline{\mathbf{f}}_{k}, \tilde{\gamma}_k; \overline{\mathbf{f}}_{k}\iLoop, \tilde{\gamma}_k\iLoop \big)
	 \geq I_k({\mathcal{B}}_k\comb) \\ 
	 &\qquad \quad \qquad \forall k, \ \forall {\mathcal{B}}_k\comb \subseteq \mathcal{B}_k, \ |{\mathcal{B}}_k\comb|\geq L_k,
	 \end{split} \label{eq:P1-KKT-SINR-2}
 \\
	&\displaystyle z_b: \ \ \ \sum_{k \in \mathcal{K}}\|\mathbf{f}_{b,k}\|^2 \leq P_b  \ \ \forall b, \label{eq:P1-KKT-Pt-2}
\end{align}
\end{subequations}
where $\mathbf{a}=[a_{1,1}, \ldots, a_{K,C(L_K)}]\tran$ and $\mathbf{z}=[z_1, \dots, z_B]\tran$ are the non-negative Lagrangian multipliers corresponding to constraints~\eqref{eq:P1-KKT-SINR-2} and~\eqref{eq:P1-KKT-Pt-2}, respectively. Dual variable $z_b$ in~\eqref{eq:P1-KKT-Pt-2} is associated with the total transmit power constraint of $b$th {RRU}. 
Dual variable $a_{k,c}$ for each user~$k$ is associated with $c$-th subset combination of SINR constraint.
%%%%and for each  user~$k$, $a_{k,c} \ \forall c$ is associated with each subset combinations of the SINR constraint. 
The role of the dual variables become clear in the following subsection.
The convex subproblem~\eqref{eq:P1-KKT-objective2} for each SCA iteration can be efficiently solved, in general, using existing convex optimization toolboxes, such as CVX~\cite{cvx}. {The fixed operating points for the current iteration are updated from the solution of the current SCA iteration. This is repeated until convergence to a stationary point.} The  beamformer design with the proposed SCA relaxation has been summarized in Algorithm~\ref{algSCA}.
\vspace{-5pt}
\SetArgSty{textnormal}
\begin{algorithm}[]
	\caption{Successive convex approximation algorithm for WSRM  problem~\eqref{eq:P1-KKT-objective2}}
	\label{algSCA}
	\SetAlgoLined
	Set $i = 1$ and initialize with a feasible starting point % choose any feasible initial points 
	$\big\{\mathbf{{f}}_{b,k}^{\scriptsize (0)}, \ \tilde{\gamma}_k^{\scriptsize (0)}\big\}$ \ \  $\forall b \in \mathcal{B}, \ \forall k \in \mathcal{K},$ \\
	\Repeat{convergence or for fixed number of iterations}{
	{Solve \eqref{eq:P1-KKT-objective2} with $\big\{\mathbf{f}_{b,k}\xiLoop, \tilde{\gamma}_k\xiLoop\big\}$ and denote the local optimal values as $\big\{\mathbf{f}_{b,k}^{*}, \tilde{\gamma}_k^{*}\big\}$ \\
	Update $\big \{\mathbf{f}_{b,k}\iLoop = \mathbf{f}_{b,k}^{*}\big \} $ and $\big \{ \tilde{\gamma}_k\iLoop = \tilde{\gamma}_k^{*} \big \}$ \\
	Set $i = i+1$
	}
}
\end{algorithm}
\vspace{-5pt}
%
%
%% ----- %%%% ----- %%
\vspace{-10px} 
\subsection{Solution via Low-Complexity KKT Conditions}
\label{subsec:KKT}
%%%%%%by parallel beamformer updates across spatailly distributed RRUs
Problem~\eqref{eq:P1-KKT-objective2} can be more efficiently solved at the BBU by the parallel update of beamforming vectors corresponding to the spatially distributed \acp{RRU}. 
Unlike~the approach presented in the previous subsection, the robust beamformer can be obtained by iteratively solving a system of \ac{KKT} equations~\cite{boyd2004convex}.
Thus, the KKT based solution provides closed-form steps for an algorithm that does not rely on generic convex solvers.
Furthermore, iterative evaluation of \ac{KKT}  optimality conditions, for each \ac{SCA} step, reveals a conveniently parallel structure for the beamformer design with significantly lower computational complexity with respect to joint beamformer optimization across all distributed RRU antennas.  
%
%, %%%%%%%%%%%%%more details on low-complexity iterative algorithm. 
%and more details are provided in following sub give a brief outline of the proposed iterative ... followed by more detailed...

The Lagrangian $\mathcal{L}(\mathbf{F}, \tilde{\gamma}_k, a_{k,c}, z_b)$ of problem~\eqref{eq:P1-KKT-objective2} is given in expression~\eqref{eq:Lagrangian-details}. 
\begin{figure*}[t!]
% ensure that we have normalsize text
\normalsize
% Store the current equation number.
\setcounter{mytempeqncnt}{\value{equation}}
% Set the equation number to one less than the one
% desired for the first equation here.
% The value here will have to changed if equations
% are added or removed prior to the place these
% equations are referenced in the main text.
\setcounter{equation}{15}
\begin{align}
\label{eq:Lagrangian-details}
%%%%%%%%%%%%%%%%%%%%%%%%%%%%%%%%%%%%%%%%%%%%
& \mathcal{L}(\mathbf{F}, \tilde{\gamma}_k, {a}_{k,c}, z_b) = - \sum\limits_{k\in \mathcal{K}} \delta_k \log \big(1+\tilde{\gamma}_k\big)
   + \sum\limits_{b \in \mathcal{B}} z_b \Big( \sum\limits_{k \in \mathcal{K}}\|\mathbf{f}_{b,k}\|^2 - P_b  \Big) 
    + \sum\limits_{k\in \mathcal{K}}  \sum\limits_{c=1}^{C(L_k)} a_{k,c} \Bigg( \sigma_k^2
    %%%\nonumber \\ & \quad
     + \sum\limits_{u \in \mathcal{K} \setminus k} \bigl| \overline{\mathbf{h}}_{k}\cherm \overline{\mathbf{f}}_{u}\bigr|^2 
   \nonumber \\ & \qquad \qquad
   - 2  \sum\limits_{j\in\mathcal{K}} \Re \bigg\{ \frac{ \overline{\mathbf{f}}_{j}\iherm \overline{\mathbf{h}}_{k}\comb \overline{\mathbf{h}}_{k}\cherm}{1 + \tilde{\gamma}_k\iLoop} \Big(\overline{\mathbf{f}}_{j} -  \overline{\mathbf{f}}_{j}\iLoop \Big) \bigg\}
   + \frac{ \sigma_k^2 + \sum\limits_{j\in \mathcal{K}} \bigl| \overline{\mathbf{h}}_{k}\cherm \overline{\mathbf{f}}_{j}\iLoop\bigr|^2}{1 + \tilde{\gamma}_k\iLoop} \bigg( 1- \frac{\tilde{\gamma}_k - \tilde{\gamma}_k\iLoop}{1 + \tilde{\gamma}_k\iLoop} \bigg)  \Bigg), \nonumber
\\ 
%%%%%%%%%%%%%%%%%%%%%%%%%%%%%%%%%%%%%%%%%%%%
%& \mathcal{L}(\mathbf{F}, \tilde{\gamma}_k, {a}_{k,c}, z_b)  = 
%   - \sum\limits_{k\in \mathcal{K}} \delta_k \log \big(1+\tilde{\gamma}_k\big) \ 
   %
%   + \sum\limits_{b \in \mathcal{B}} z_b \bigg( \sum\limits_{k \in \mathcal{K}}\|\mathbf{f}_{b,k}\|^2 - P_b  \bigg) %
   %
%   + \sum\limits_{k\in \mathcal{K}} \sum\limits_{c\in \mathcal{C}_k}  a_{k,c} \Bigg( 
   %
   %%%%%\sigma_k^2 + \sum_{u \in \mathcal{K} \setminus  k} \Bigl| \sum_{b\in {\mathcal{B}}_u \setminus \mathcal{D}_k\comb}\mathbf{h}_{b,k}\herm \mathbf{f}_{b,u}\Bigr|^2 \nonumber  \\ & \qquad \qquad \qquad \qquad \qquad \qquad
   %
%   \sigma_k^2 + \sum\limits_{u \in \mathcal{K} \setminus  k} \mid \overline{\mathbf{h}}_{k}\cherm \overline{\mathbf{f}}_{u}\mid ^2 \nonumber  \\ & \qquad \qquad \qquad
   %
%   - 2  \sum\limits_{j\in\mathcal{K}} \Re \bigg\{ \frac{ \overline{\mathbf{f}}_{j}^{(i){\mbox{\scriptsize H}}} \overline{\mathbf{h}}_{k}\comb \overline{\mathbf{h}}_{k}\cherm}{1 + \tilde{\gamma}_k^{(i)}} \Big(\overline{\mathbf{f}}_{j} -  \overline{\mathbf{f}}_{j}^{(i)} \Big) \bigg\} \ - 
   %
%   \frac{ \sigma_k^2 + \sum\limits_{j\in \mathcal{K}} \bigl| \overline{\mathbf{h}}_{k}\cherm \overline{\mathbf{f}}_{j}\iLoop\bigr| ^2}{1 + \tilde{\gamma}_k^{(i)}} \bigg( 1- \frac{\tilde{\gamma}_k - \tilde{\gamma}_k^{(i)}}{1 + \tilde{\gamma}_k^{(i)}} \bigg)  \Bigg). \\
    %%%%%%%%%%%%%%%%%%%%%%%%%%
%%%%& \mathcal{L}(\mathbf{F}, \tilde{\gamma}_k, {a}_{k,c}, z_b) 
%
& \quad = - \sum\limits_{k\in \mathcal{K}} \delta_k \log \big(1+\tilde{\gamma}_k\big) 
   + \sum\limits_{b \in \mathcal{B}} z_b \Big( \sum\limits_{k \in \mathcal{K}}\|\mathbf{f}_{b,k}\|^2 - P_b  \Big) 
   + \sum\limits_{k\in \mathcal{K}} \bigg( \sum\limits_{c=1}^{C(L_k)}  a_{k,c} \sigma_k^2 
   +  \sum\limits_{u \in \mathcal{K} \setminus  k}  \sum\limits_{c=1}^{C(L_u)} a_{u,c} \bigl| \overline{\mathbf{h}}_{u}\cherm \overline{\mathbf{f}}_{k} \bigr|^2 \nonumber \bigg) 
   \nonumber \\ & \qquad \qquad
   - \sum\limits_{k\in \mathcal{K}} \Bigg( 2 \sum\limits_{j\in\mathcal{K}} \sum\limits_{c=1}^{C(L_j)} 
   a_{j,c} \Re \bigg\{ \frac{ \overline{\mathbf{f}}_{k}\iherm \overline{\mathbf{h}}_{j}\comb \overline{\mathbf{h}}_{j}\cherm}{1 + \tilde{\gamma}_j\iLoop}
    \Big(\overline{\mathbf{f}}_{k} -  \overline{\mathbf{f}}_{k}\iLoop \Big) \bigg\} 
    + \sum\limits_{c=1}^{C(L_k)}  a_{k,c}
    \frac{ \sigma_k^2 + \sum_{j\in \mathcal{K}} \bigl| \overline{\mathbf{h}}_{k}\cherm \overline{\mathbf{f}}_{j}\iLoop \bigr|^2}{1 + \tilde{\gamma}_k\iLoop}
    \bigg( 1- \frac{\tilde{\gamma}_k - \tilde{\gamma}_k\iLoop}{1 + \tilde{\gamma}_k\iLoop} \bigg) \Bigg) .
\end{align}
% Restore the current equation number.
%%\setcounter{equation}{\value{mytempeqncnt}}
\setcounter{equation}{16}
% IEEE uses as a separator
\hrulefill
% The spacer can be tweaked to stop underfull vboxes.
\vspace*{-15pt}
\end{figure*}
%
%%%% KKKT conditions for the centralized case in Appendix
%
%%%%%%%%%%%%It should be noted that the solution which attains the maximum of Lagrangian is also primal solution, if and only if, such solution is both primal feasible and unique~\cite{boyd2004convex}. Therefore, it is sufficient to show that the approximated convex subproblem satisfies the Slater's conditions~\cite{boyd2004convex}. 
%
It should be noted that the \ac{KKT} optimality conditions provide necessary and sufficient conditions for the solution of a convex problem~\cite[Ch. 5]{boyd2004convex}. 
%
%%%%%%%%%%%Therefore, it is sufficient to show that the approximated convex subproblem~\eqref{eq:P1-KKT-objective2} satisfies the Slater's conditions~\cite{boyd2004convex, bertsekas2003convex}.
%
Thus, the solution for problem~\eqref{eq:P1-KKT-objective2} can be obtained by iteratively solving a system of \ac{KKT} optimality conditions, which include stationary, complementary slackness, and primal-dual feasibility requirements (for more detailed derivation see Appendix~\ref{app:KKT-Conditions}). Next, we briefly outline the key challenges in solving the problem~\eqref{eq:P1-KKT-objective2} by using \ac{KKT} conditions and our proposed solution.
%
%
%%%%practical challenge 1: overlapping interdependent SINR constraints with individual dual variables, solved via subgradient
%
%%%It is worth pointing out that in~\eqref{eq:P1-KKT-SINR-2}, 

The user-specific \ac{SINR} constraints~\eqref{eq:P1-KKT-SINR-2} are mutually interdependent over the link blockage combinations (see Section~\ref{subsec:blockage}). Thus, it makes deriving an efficient and closed-form solution 
%%%%via the \ac{KKT} optimality conditions
for Lagrangian multipliers $a_{k,c} \ \forall(k,c)$ considerably more difficult than in the case with only a single \ac{SINR} constraint per-user~\cite{ Kaleva-DecentralizeSumRate-2016, Ganesh-TrafficAware-2016, Tervo-EE-2017, Kaleva-DecentralizeJP-2018, Tervo-EE-2018}. 
%%%%%To overcome this, Lagrangian multipliers $a_{k,c} \ \forall(k,c)$ corresponding to each coupled SINR constraint are iteratively solved using the subgradient method.
To overcome this, we resort to a subgradient approach, where Lagrangian multipliers $a_{k,c} \ \forall(k,c)$  are  solved using the constrained ellipsoid method~\cite{boyd2003subgradient}.

%corresponding to coupled SINR constraint
%%>>>>>practical challenge 2: BS specific subvectors of the joint beamforming vector are subject BS specific power constraints with individual dual variables... -> solution  is to parallelize the process using the best response...
%
%Another practical challenge is that 
In addition, the design of optimal beamformer $\mathbf{F}$ is inherently coupled between all distributed RRU antennas due to coherent joint transmission to each user.   
%%%%%>>>>>global joint beamformer solved from Lagrangian directly - large matrix to be inverted + dual variables $z_b$ challenging to find jointly to meet the power constraints. Solution
Therefore, the computational complexity of optimal beamformer $\overline{\mathbf{f}}_k \ \forall k$ scales cubically with the length of joint beamformers~$(BN_t)$, which quickly becomes intractable, e.g., for dense deployments. Furthermore, RRU specific dual variables $z_b \ \forall b$ should be computed simultaneously~(see \eqref{eq:Derivate_CompletePrecoder} in Appendix~\ref{app:KKT-Conditions}). However, because of coupling and interdependence among Lagrangian multipliers $z_b \ \forall b$ due to \ac{JT}-\ac{CoMP}, it is computationally challenging to obtain a closed-form solution.  %%their exact values. 
To overcome this, we also incorporate a parallel optimization framework using the best response~\cite{Kaleva-DecentralizeJP-2018} into the iterative optimization process (see~\eqref{eq:Derivate_Precoder} in Appendix~\ref{app:KKT-Conditions}). As a result, RRU specific beamformers are solved in parallel for each iteration, while assuming that coupling from other cooperating RRUs is fixed to the solution from the previous iteration. Note that for a convex problem, monotonic convergence can be guaranteed by a regularization step on beamformer update~\cite{Scutari-2014-Decomposition}.  

In the following, we provide an iterative algorithm by combining the SCA framework with dual and best response methods, which admits the closed-form solution in each step. Specifically, in each SCA iteration, the approximated convex subproblem~\eqref{eq:P1-KKT-objective2} is solved  via the iterative evaluation of the \ac{KKT} optimality conditions. {Furthermore, to improve the rate of convergence, the SCA approximation point $\big\{\mathbf{{f}}_{b,k}\iLoop, \tilde{\gamma}_k\iLoop \big\}$ is also updated in each iteration along with the Lagrangian multipliers} (for more detailed derivation see Appendix~\ref{app:KKT-Conditions}).

To summarize, the steps in the iterative algorithm are
\begin{subequations}
\label{eq:KKT-i iteration}
\begin{align}
& \mathbf{f}_{b,k}\Starherm  =  
\Big( \mathbb{I}z_b + \sum\limits_{u \in \mathcal{K} \setminus  k} \sum\limits_{c=1}^{C(L_u)}  a_{u,c}\xiLoop \mathbf{h}_{b, u}\comb \mathbf{h}_{b, u}\cherm \Big)^{-1}  \mathbf{t}_{b,k}\xiLoop , \label{eq:KKT-i, f*} \\
& \mathbf{{f}}_{b,k}\iLoop = \mathbf{{f}}_{b,k}\xiLoop + \psi \Big[ \mathbf{{f}}_{b,k}^{*} - \mathbf{f}_{b,k}\xiLoop \Big] ,
\label{eq:KKT-i, f i}  \\
& \Gamma_k\iLoop({\mathcal{B}}_k\comb) = %%\frac{ \Bigl|\sum\limits_{b\in {\mathcal{B}}_k\comb}\mathbf{h}_{b,k}\herm \mathbf{f}_{b,k}\iLoop \Bigr|^2  }{\sigma_k^2 + \sum\limits_{u \in \mathcal{K} \setminus  k} \Bigl| \sum\limits_{b\in {\mathcal{B}}_u \setminus \mathcal{D}_k\comb}\mathbf{h}_{b,k}\herm \mathbf{f}_{b,u}\iLoop \Bigr|^2},
\frac{ | \overline{\mathbf{h}}_{k}\cherm \overline{\mathbf{f}}_{k}\iLoop |^2  }{\sigma_k^2 + \sum\limits_{u \in \mathcal{K} \setminus  k} \bigl| \overline{\mathbf{h}}_{k}\cherm \overline{\mathbf{f}}_{u}\iLoop\bigr|^2} ,
\label{eq:KKT-i, SINR i} \\
& \tilde{\gamma}_k\iLoop ={\delta_k} \Bigg\{ \sum\limits_{c=1}^{C(L_k)} a_{k,c}\xiLoop \frac{ \sigma_k^2 + \sum\limits_{j\in \mathcal{K}} \bigr| \overline{\mathbf{h}}_{k}\cherm \overline{\mathbf{f}}_{j}\xiLoop \bigl| ^2}{\big(1 + \tilde{\gamma}_k\xiLoop\big)^2} \Bigg\}^{-1} - 1 ,
\label{eq:KKT-i, gamma i} \\
& a_{k,c}\iLoop = \bigg( a_{k,c}\xiLoop + \beta \Big[ \tilde{\gamma}_k\iLoop - \Gamma_k\iLoop({\mathcal{B}}_k\comb) \Big] \bigg)^{+} ,
\label{eq: KKT-i ak}
\end{align}
\end{subequations}
where 
\begin{align*}
\mathbf{t}_{b,k}\xiLoop = &
 \sum\limits_{j\in\mathcal{K}} \sum\limits_{c=1}^{C(L_j)} a_{j,c}\xiLoop \frac{\Big(\overline{\mathbf{f}}_{k}\xiherm \overline{\mathbf{h}}_{j}\comb \Big)}{1+\tilde{\gamma}_j\xiLoop} \mathbf{h}_{b,j}\cherm \nonumber \\ & \ \ 
- \sum\limits_{u \in \mathcal{K} \setminus  k } \sum\limits_{c=1}^{C(L_u)}  a_{u,c}\xiLoop \bigg( \sum_{g \in \mathcal{B} \setminus  b } \mathbf{f}_{g,k}\xiherm \mathbf{h}_{g,u}\comb \mathbf{h}_{b,u}\cherm \bigg) ,
\end{align*} 
and $\mathbf{h}_{b,k}\comb = \mathbbm{1}_{\mathcal{G}_k\comb}(b)\mathbf{h}_{b,k}$ for all $c=1,\dots,C(L_k)$, $k \in \mathcal{K}$, {and} $b \in \mathcal{B}$. The best response and subgradient step sizes are represented with $\psi>0$ and $\beta>0$, respectively. In~\eqref{eq: KKT-i ak}, we have use $(x)^{+}\triangleq\mathrm{max}(0, x)$. %%%% are small positive step sizes, and $(x)^{+}\triangleq\text{max}(0, x)$. 
The expressions in~\eqref{eq:KKT-i iteration} are solved in an iterative manner, starting with initializing the variables $\{\mathbf{{f}}_{b,k}^{\scriptsize (0)}, \ \tilde{\gamma}_k^{\scriptsize (0)}, \ a_{k,c}^{\scriptsize (0)}\}$ with feasible values, such that SINR and total transmit power constraints for each distributed RRU are satisfied. 
Note that due to reformulation of constraint~\eqref{eq:P1-C1} as in~\eqref{eq:Compact-SINR+1}, we get $\{1+\tilde{{\gamma}}_j\}_{j\in\mathcal{K}}$ in the denominator of~\eqref{eq:KKT-i, f*} and~\eqref{eq:KKT-i, gamma i}, and these are invertible, even if subset of users are assigned zero SINR. Thus, the proposed iterative method for problem~\eqref{eq:P1-KKT-objective2} is numerically stable. %%, and it is summarized in Algorithm~\ref{algKKT}. 
The beamformer design by iteratively solving a system of \ac{KKT} optimality conditions is summarized in Algorithm~\ref{algKKT}.
%
%% ----- %%%% ----- %%
\vspace{-5pt}
\SetArgSty{textnormal}
\begin{algorithm}[]
	\caption{Low-complexity KKT based iterative algorithm for {WSRM} problem~\eqref{eq:P1-KKT-objective2}}
	\label{algKKT}
	\SetAlgoLined
	Set $i = 1$ and initialize with a feasible starting point %choose any feasible initial points 
	$\big\{\mathbf{{f}}_{b,k}^{\scriptsize (0)}, \tilde{\gamma}_k^{\scriptsize (0)}, a_{k,c}^{\scriptsize (0)}\big\}  \ \forall b \in \mathcal{B}, \forall k \in \mathcal{K}, \forall c=1,\dots,C(L_k)$,  \\ 
	\Repeat{convergence or for fixed number of iterations}{
	{
	Solve $\mathbf{f}_{b,k}^{*}$ from~\eqref{eq:KKT-i, f*} with $\big\{\mathbf{{f}}_{b,k}\xiLoop, \tilde{\gamma}_k\xiLoop, a_{k,c}\xiLoop\big\}$ \\ 
	Update $\mathbf{f}_{b,k}\iLoop$ using~\eqref{eq:KKT-i, f i} \\
	Calculate  $\tilde{\gamma}_k\iLoop$ from~\eqref{eq:KKT-i, gamma i}  \\
	Update $a_{k,c}\iLoop$ using~\eqref{eq: KKT-i ak} with~$\big\{\tilde{\gamma}_k\iLoop, \Gamma_k\iLoop({\mathcal{B}}_k\comb)\big\}$ \\
	Set $i = i+1$
	}
}
\end{algorithm}
\vspace{-5pt}
%% ----- %%%% ----- %%
%

\subsubsection{Lagrangian Multipliers}
\label{subsubsec:coupledSINR}
%%%% Explanation of a_(k,c)
Dual variables $a_{k,c} \ \forall (k,c)$ corresponding to constraint~\eqref{eq:P1-KKT-SINR-2} are interdependent and mutually coupled due to the common SINR constraint across the link blockage combinations. Their exact values for each SCA iteration can not be obtained as a closed-form expression. Therefore, we resort to a widely used subgradient approach, such as the constrained ellipsoid method, which converges to the local optimal solution for a convex optimization problem~\cite{boyd2003subgradient}. It should be noted that choice of the step size $\beta$ in expression~\eqref{eq: KKT-i ak} depends on the system model, as it directly affects the convergence rate as well as control the oscillation in the \ac{WSRM} objective function. There have been several studies in the literature on the convergence properties of the subgradient approach, with the different step size rules \cite{boyd2003subgradient, bertsekas2003convex}. More precisely, monotonic convergence can not be guaranteed, in general, for the constrained ellipsoid method, and thus, one has to track and adjust the step size accordingly. In the proposed iterative approach in Algorithm~\ref{algKKT}, the dual variables $a_{k,c} \ \forall (k,c)$ are updated based on the violation of the \ac{SINR} constraint with a small positive step size, as in~\eqref{eq: KKT-i ak}. 

%%%% Explanation of cB
Dual variables $z_b \ \forall b$ are chosen to satisfy the transmit power constraint~\eqref{eq:P1-KKT-Pt-2}, using the bisection search method. It should be noted, in a multi-cell scenario, the transmit power constraints may not necessarily always hold with equality. Specifically, for each RRU $b$, if $\sum_{k \in \mathcal{K}}\|\mathbf{{f}}_{b,k}^{*}\|^2 < P_b $ then $z_b = 0$, i.e., non-negative dual variable $z_b$ is set to zero in order to satisfy the corresponding complementary slackness conditions~\cite[Ch.~5.5.2]{boyd2004convex}. Otherwise, there exist a unique $z_b>0$ such that $\sum_{k \in \mathcal{K}}\|\mathbf{{f}}_{b,k}^{*}\|^2 = P_b$ for all $b\in\mathcal{B}$. %%%This value is unique and provides the  solution~\cite{boyd2004convex}, which can be calculated by  the bisection search with respect to the total power constraint.

\subsubsection{Best Response}
\label{subsubsec:Best Response}
The  beamformer $\mathbf{F}$ is inherently coupled among all distributed RRU antennas~\eqref{eq:KKT-i iteration}, because of the coherent joint transmission to each user.
One possible approach is based on updating the beamformers sequentially, i.e., using the Gauss-Seidel type update process, which provides monotonic convergence for a \ac{WSRM} optimization problems. However, it is shown in~\cite{kaleva-2014-decentralized} that the convergence rate drastically reduces even with a slight increase in the number of cooperating RRUs.  
Here, instead, we implement a parallel optimization framework~\cite{Scutari-2014-Decomposition}, which efficiently parallelizes the beamformer updates across the distributed \ac{RRU} antennas, and hence significantly reduces the per-iteration computational complexity. Specifically, for a given iteration, RRU specific beamformers are solved in parallel while assuming the coupling from other RRUs is fixed to the solution from the previous iteration, as in expression~\eqref{eq:KKT-i, f*}.
{The objective function can be shown to converge if we allow a sufficiently large number of subgradient iterations per fixed SCA approximation (until increased objective) for each \ac{RRU} before making the best response step  with a sufficiently small step size~\cite{Kaleva-DecentralizeJP-2018}. However, here we are more interested in a fast and robust rate of convergence, for which, we allow only a single subgradient update per best response iteration. Thus, the Algorithm~\ref{algKKT} may not converge to the same point as Algorithm~\ref{algSCA}.} It is shown by numerical examples in Section~\ref{sec:Sim-Result} that this still provides excellent performance with a fairly small number of iterations.
More details on the convergence behavior and choice of step size $\psi\in(0,1)$ with the best response based parallel optimization approach 
are provided in~\cite{Scutari-2014-Decomposition}. %%%For the time invariant channels, convergent~$\psi$ can be bounded to Lipschitz constant of the objective~\cite{Scutari-2014-Decomposition}.
It should be noted that the RRU-specific transmit power constraints are convex (see \eqref{eq:P1-KKT-Pt-2}), therefore, regularized update with $\psi < 1$ in~\eqref{eq:KKT-i, f i} will strictly preserve the feasibility of total transmit power constraint of each RRU.

\subsubsection{Feasible Initial Point}
\label{subsubsec:Feasible Initial Point}
In the \ac{SCA} framework, all non-convex constraints are approximated with a sequence of convex subsets and then iteratively solved until convergence of the objective~\cite{marks1978general, beck2010sequential}. Thus, it is very important to initialize the iterative algorithm with some feasible starting point.
To this end, one possible solution for the feasible initial $\mathbf{f}_{b,k}^{\scriptsize(0)}$ is to use any beamformer satisfying the transmit power constraint~\eqref{eq:P1-C2}, which can  be obtained by scaling a randomly generated beamforming vector. Then, the lower bound on achievable SINR can be calculated from expression~\eqref{eq:SINR-comb}, i.e., $\tilde{\gamma}_k^{\scriptsize (0)}=\underset{c}{\mathrm{min}}\big( \Gamma_k({\mathcal{B}}_{k}\comb) \big)$ $\forall c=1,\dots,C(L_k)$.
However, it should be noted that the randomly generated initial solution can be very far from the optimal solution and may require a significantly large number of iterations until convergence.
As an example, for a system model with $N_t \geq K$, an efficient initial point can be obtained by simply matching the beamformers $\mathbf{f}_{b,k}^{\scriptsize(0)}$ to the corresponding channel $\mathbf{h}_{b,k} \ \forall (b,k)$, i.e., based on \ac{MRT}, while neglecting the potential intra-cell and inter-cell interference. In addition, non-negative dual variables $a_{k,c}^{\scriptsize(0)} \ \forall (k,c)$ are randomly chosen such that left-hand-side (LHS) of expression~\eqref{eq:KKT-i, gamma i} is strictly non-negative (see~\eqref{eq:aK_sum-nonZero} in Appendix~\ref{app:KKT-Conditions}). 
It should be noted that initializing the algorithm with different feasible initial values does not, on the average, affect the final  solution of problem~\eqref{eq:P1-KKT-objective2} given a sufficient number of iterations. 
%
%%%%%\revise{We refer the reader to \cite[Appendix A]{Ganesh-TrafficAware-2016} for the details on the convergence properties of the \ac{SCA} framework based non-convex SINR relaxation to the stationary point.}
{For more details on the stationary point solution and the convergence properties of SCA framework, we refer the reader to~\cite[Appendix A]{Ganesh-TrafficAware-2016}.}

\subsubsection{Complexity Analysis}
\label{subsubsec:Complexity Analysis}
The approximated convex subproblem~\eqref{eq:P1-KKT-objective2} can be solved in a generic convex optimization solver as a sequence of second-order cone programs (SOCP)~\cite{Lobo-Vandenberghe-Boyd-Lebret-98}. The~complexity of the  problem scales exponentially with the length of the joint beamformers $(BN_t)$ and the number of  constraints~\cite{Lobo-Vandenberghe-Boyd-Lebret-98}. 
Thus, particularly, for dense mmWave deployments with large $N_t$ and $B$, the complexity quickly becomes intractable in practice. 
The complexity of the proposed KKT based iterative solution is dominated by~\eqref{eq:KKT-i, f*}, which mainly consists of matrix multiplications and inverse operations and that scale with \ac{RRU} specific beamformer size~$(N_t)$. %%%%Furthermore, the dimensions of all inverse operation in~\eqref{eq:KKT-i iteration} are directly proportional to number of transmit antennas $(N_t)$ at each RRU. 
{Therefore, the worst-case computational requirement is $\mathcal{O}\big({\kappa\tau}|B_k|N_t^3\big)$, where $\kappa$~is the number of iterations, $\tau$ is the number of bisection steps per iteration, $|B_k|$ number of \acp{RRU} that serve $k$th user, and $N_t$ is digital beamformer size, e.g., the number of antennas at the \ac{RRU}.}
In addition, the complexity of matrix inversion operation in~\eqref{eq:KKT-i, f*} can be alleviated by solving $\mathbf{f}_{b,k}^{*}$ from a system of linear equations, providing a significant reduction in the computational complexity. %%%%. Moreover, the implementation complexity is fairly small due to the closed-form arithmetic steps~\eqref{eq:KKT-i iteration}. %%
  %%%for even modestly sized systems. 
%
%%%%%%%\revise{Hence, despite its iterative nature, the Algorithm~\ref{algKKT} brings a substantial computational complexity reduction and the implementation complexity is small due to the closed-form arithmetic representations of the steps in~\eqref{eq:KKT-i iteration}.}
%

{In the proposed methods, the centralized BBU computes the beamformers $\mathbf{f}_{b,k}$ $\forall b\in\mathcal{B}$, $k\in\mathcal{K}$ based on the available CSI. Hence, there is no additional signalling exchange and overhead among the cooperating RRUs. Therefore, the signalling overhead of the proposed algorithms is admissible and supported by the C-RAN architecture in the upcoming $5$G systems~\cite{CRAN-2015}.}

\textit{Extensions}: The iterative evaluation of \ac{KKT} conditions can be extended to \ac{ADMM} design, wherein mutually coupled and non-convex \ac{SINR} constraint~\eqref{eq:P1-C1} can be handled by augmented Lagrangian method. In addition, problem~\eqref{eq:P1-objective} can be further extended by tuning the parameters  $\{\mathcal{B}_k,  L_k\}$ for all $k \in \mathcal{K}$ based on statistical information. The \ac{ADMM} design and joint optimization of parameters~$\{\mathcal{B}_k, L_k\}_{k\in\mathcal{K}}$ are left for future work.

\setcounter{equation}{17}
%% --- Simulations Results --------------------------
%\vspace{-10px}
\section{Simulation Results}
\label{sec:Sim-Result}
%
%\vspace{-5px}
%
This section provides numerical results to validate the performance of the proposed methods. In particular, we analyze the impact of subset size $L_k$ for all $k$ on outage performance and  sum-rate, as well as evaluate the trade-off between these performance metrics. We further elaborate on the convergence and the performance gap between the proposed algorithms.     

%\vspace{-10px}
%
\subsection{Simulation Setup}
\label{subsec:sim-setup}
\begin{figure}[t]
\setlength\abovecaptionskip{-0.25\baselineskip}
\centering
\includegraphics[trim=0.3cm 0.1cm 0.3cm 0.3cm, clip, width=0.95\linewidth]{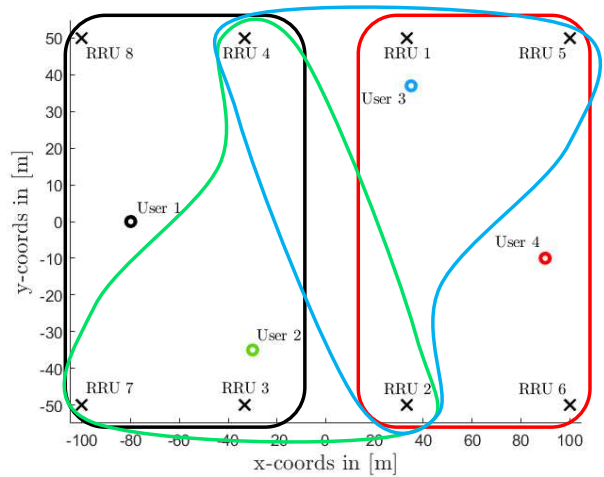}
\caption{Downlink system model showing user-centric RRU-user pairing  for a given channel realization with $|\mathcal{B}|=8$ and $|\mathcal{B}_k|=4 \ \forall k$.}
\label{fig:RRU-user_pair}
\end{figure}

%
%%%%%%%%%%%%%\revise{We consider a mmWave based downlink system with \revise{$B = 8$ RRUs} and each \ac{RRU} is equipped with \acp{ULA} of $N_t = 16$ antennas\footnote{\revise{The proposed algorithms presented in the sequel of the paper can be extended to any antenna array geometry.}}. All RRUs are placed in a rectangle layout and connected to common BBU in the edge cloud. It not mentioned otherwise, we consider \ac{JT}-\ac{CoMP} scenario with partially overlapping user-centric clusters, i.e., $ |\mathcal{B}_k| = 4 \ \forall k$, such that all $K = 4$ users are coherently served by its serving set, e.g., $\mathcal{B}_k (\subseteq \mathcal{B})$ RRUs. As an example, for a given channel realization, the RRU-user pairing with partially overlapping user-centric clusters is shown in~Fig.~\ref{fig:RRU-user_pair}.  %% (excluding the potentially blocked~links).  
%
%%%%%\footnote{\revise{The proposed algorithms presented in the sequel of the paper can be extended to any antenna array geometry.}}
%
{We consider a mmWave based downlink system with $B=8$~RRUs and each RRU is equipped with \acp{ULA} of $N_t=16$ antennas. All RRUs are placed in a rectangle layout (resembling e.g., a factory-type setup) and connected to a common BBU in the edge cloud. If not mentioned otherwise, we consider \ac{JT}-\ac{CoMP} scenario with partially overlapping user-centric clusters, such that each active user~$k$ receives a coherently synchronous signal from all RRUs in $\mathcal{B}_k (\subseteq \mathcal{B})$, as shown in Fig.~\ref{fig:RRU-user_pair} for a given channel realization. 
Recall that to improve communication reliability, we use parameter $L_k(\leq|\mathcal{B}_k|)$\footnote{{The joint optimization of design parameters parameters~$\{\mathcal{B}_k, L_k\}_{k\in\mathcal{K}}$  is an interesting topic for future studies.}} and proactively model the \ac{SINR} over the link blockage combinations (see Section~\ref{sec:ReliablityVSCoMP}). 
For simplicity, let us assume $ |\mathcal{B}_k|=4$, $L_k = L$ for all $k\in\mathcal{K}$. All~RRUs are assumed with the same maximum transmit power, i.e., $P_b~=~33~\text{dBm}$ for all $b\in\mathcal{B}$. The AWGN noise is set to $-72$~dBm/Hz, carrier-frequency $f_c = 28$~GHz, and a $20$~MHz frequency band is assumed to be fully reused across all RRUs. In the simulations, we set LoS path-loss exponent $\varrho=2$, NLoS path-loss exponent $\zeta\in[2, \ 6]$, and the user priorities are set to be equal (i.e., $\delta_k=1 \ \forall k$).
%
%
%%%%%%%%%%%%%%%%For simplicity, let $L_k = L \ \forall k$, all RRUs are assumed with the same maximum transmit power, i.e., $P_b=P_t \ \forall b$, carrier-frequency $f_c = 28$~GHz and cell-edge $\text{SNR}=15$~dB. More precisely, \revise{$\mathrm{SNR}=(P_{t}/\sigma_k^2) d_{e}^{-{\varrho}} \ \forall b$,  where $\sigma_k^2$ is noise variance, ~$d_{e}$ denotes the cell-edge distance and $\varrho=2$ is LoS path-loss exponent.} The AWGN noise is set to $-78$~dBm/Hz and a $20$~MHz frequency band is assumed to be fully reused across all RRUs. In the simulation, we set NLoS path-loss exponent $\zeta\in[2, \ 6]$. The user priorities are set to be equal (i.e., $\delta_k=1 \ \forall k$).
}
All $K=4$ users are assumed to be randomly dropped within the coverage region, hence having different path gain and angle with respect to each RRU.
By default, each antenna is equipped with a dedicated RF chain and data converters to enable fully digital signal processing. Hybrid analog-digital beamforming structures are evaluated in Section~\ref{subsec:Hybrid}. In expression~\eqref{eq:KKT-i iteration}, we use $\beta=0.005$ and~$\psi = 0.05$  for the subgradient and best response step sizes, respectively. 
%
%
%%For the baseline methods, we consider \ac{MRT}, \ac{CB}~\cite{Antti-OntheValue-2009} and full-JT~\cite{Kaleva-DecentralizeJP-2018} beamformer design.  
%

%
%

%% ----- %%%% ----- %%
%
%\vspace{-10px}
\subsection{Outage Performance}
\label{subsec:otage-Results}
\begin{figure}[t]
\setlength\abovecaptionskip{-0.25\baselineskip}
\centering
\includegraphics[trim=0.25cm 0.08cm 0.25cm 0.2cm, clip, width=0.95\linewidth]{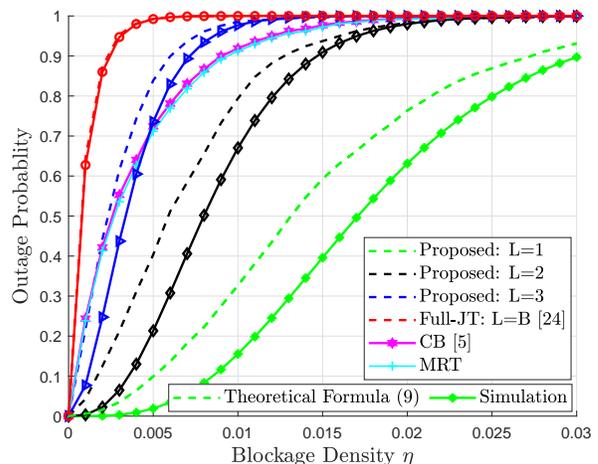}
\caption{Outage as a function of increasing blockage density with theoretical formula~\eqref{eq:Pout_Formula} (dashed line) and simulations (solid~line).}
\label{fig:outage-Results}
\end{figure}

Fig.~\ref{fig:outage-Results} shows the outage performance as a function of increasing blockage density~$\eta$ for the \ac{WSRM} problem.
%
%
%%\begin{figure}[t]
%%\centering
%%\includegraphics[trim=0.2cm 0.1cm 0.2cm 0.1cm, clip, width=\linewidth]{figs/ProbOutageMar2.eps}
%%\caption{Outage performance as function of increasing link blockage probability with theoretical formula~\eqref{eq:Pout_Formula} (solid line) and simulations (dotted line).}
%%\label{fig:outage-ResultsProb}
%%\end{figure}
%
%%%%%%%trim=0.25cm 0.1cm 0.25cm 0.1cm, clip, 
Outage event occurs if the assigned transmit rate~$R_k$ exceeds the achievable link rate\footnote{{For a given parameter} $L$, let us denote solution $\mathcal{S}_k=\{ \tilde{\gamma}_k^{*}, \overline{\mathbf{f}}_k^{*} \}_{k\in\mathcal{K}}$ obtained from the Algorithm~\eqref{algKKT}. Then, for each user~$k$, the transmission rate $R_k=\log_2(1+\tilde{\gamma}_k^{*})$ (see Section~\ref{sec:ReliablityVSCoMP}). 
However, the supported link rate of each user $k$ depends on the obtained beamformers $\{\overline{\mathbf{f}}_k^{*}\}_{k\in\mathcal{K}}$ and current channel state $\{\mathbf{h}_{b,k}\}_{b\in \mathcal{B}, k \in\mathcal{K}}$, which can not be exactly known to the \ac{BBU} during data transmission phase due to random blockages (e.g., due to channel aging of blockage effect). Hence, the supported rates are calculated using the actual \ac{SINR} values~\eqref{eq:SINR}, i.e., $C_k=\log_2( 1+\Gamma_k(\mathcal{B}_k) ) \ \forall k$, and these rates are unknown to the \ac{BBU}.
%
%%%%However, the supported link rate $C_k$ using the beamformer $\overline{\mathbf{f}}_{k}$ depends on the actual \ac{SINR} values~(see expression~\eqref{eq:SINR}), i.e., $C_k=\delta_k\log_2( 1+\Gamma_k(\mathcal{B}_k) ) \ \forall k$, and these rates are unknown to \ac{BBU} due to random link blockages.
}~$C_k$, for any user~$k=1,\ldots,K$, i.e., 
\begin{equation}
\label{eq:outage-def-sim}
P_{\mathrm{out}} \triangleq \mathbf{P} \big\{ R_k > C_k \big\} \ \ \forall k \in \mathcal{K}.
\end{equation}
It can be concluded from Fig.~\ref{fig:outage-Results} that the outage performance is greatly improved by decreasing the subset size $L$. Clearly, lower $L$ provides more stable and robust communication. 
The beamformer design with $L=1$ can provide reliable connectivity even if all but one LoS links are blocked. For example, with the blockage density $\eta = 0.005$, the outage probability is decreased from $99\%$ to less than $5\%$ by changing the parameter $L$ from $4$~to~$1$ in problem~\eqref{eq:P1-KKT-objective2}. 
%
%
%%%%%In Fig.~\ref{fig:outage-ResultsProb} we assume a binary probabilistic blockage model~\cite{Dileep_Globecom2019}. Specifically, a LoS link between RRU-user pair is blocked, i.e.,~$\mathbf{h}_{b,k}^{\text{LoS}} = \mathbf{0}$, with the probability of $q_{b,k}~\in~[0,1]$, for all $b~\in~\mathcal{B}$ and $\ k \in \mathcal{K}$. 
%
%%%%%%Similar to Fig.~\ref{fig:outage-Results}, we can conclude from Fig.~\ref{fig:outage-ResultsProb} 
Thus, the specific rate allocation with $L=1$ can withstand blockage up to a single active LoS link, and provides greatly improved communication reliability.

When comparing the simulated results with the theoretical approximation~\eqref{eq:Pout_Formula}, for smaller values of $L$, the simulated outage performance appears relatively better than the theoretical counterpart, as the \ac{WSRM}  problem~\eqref{eq:P1-KKT-objective2} solved at the BBU may end up assigning non-zero powers to only a subset of users, while all remaining users are assigned zero rate. In such a scenario, missing a LoS link results in a different blockage than what is predicted by the theoretical formula~\eqref{eq:Pout_Formula}. Moreover, the expression~\eqref{eq:Pout_Formula} models the case when the channel between a RRU-user pair is either fully-available or completely blocked (e.g., both LoS and NLoS paths in~\eqref{eq:channel-def}).
However, an increase in the subset size $L$ also increases the SINR estimate (see~\eqref{eq:SINR-comb}). Thus, it is likely that all users are assigned with some non-zero downlink rate, and, therefore, the simulated results tend to closely match to theoretical results obtained from~\eqref{eq:Pout_Formula}.
Furthermore, it can be seen from Fig.~\ref{fig:outage-Results} that the proposed methods significantly outperform the conventional full-\ac{JT} ($L=B$), \ac{CB} and \ac{MRT} based beamformer designs.

%% ----- %%%% ----- %%
%\vspace{-10px}
\subsection{Effective Sum-rate Performance}
\label{subsec:sum-rate-Results}
%
%
\begin{comment}
\begin{figure}[t]
%%\setlength\abovecaptionskip{-0.2\baselineskip}
\centering
\includegraphics[trim=0.25cm 0.08cm 0.25cm 0.2cm, clip, width=0.5\linewidth]{figs/sumRate_Review.eps}
\caption{Effective sum-rate as a function increasing blockage density.}
\label{fig:throughput-Results}
\end{figure}
%
%
\begin{figure}[t]
%%\setlength\abovecaptionskip{-0.2\baselineskip}
\centering
\includegraphics[trim=0.25cm 0.08cm 0.25cm 0.2cm, clip, width=0.5\linewidth]{figs/SumRateCurve_Review.eps}
\caption{Effective sum-rate as a function different choice of parameter $L$.}
\label{fig:throughput-L-Res}
\end{figure}
\end{comment}
%
%
\begin{figure}[t]
%%\begin{minipage}[c]{0.49\textwidth}
\setlength\abovecaptionskip{-0.25\baselineskip}
\centering
\setlength\abovecaptionskip{-0.25\baselineskip}
\includegraphics[trim=0.25cm 0.08cm 0.25cm 0.2cm, clip, width=0.95\linewidth]{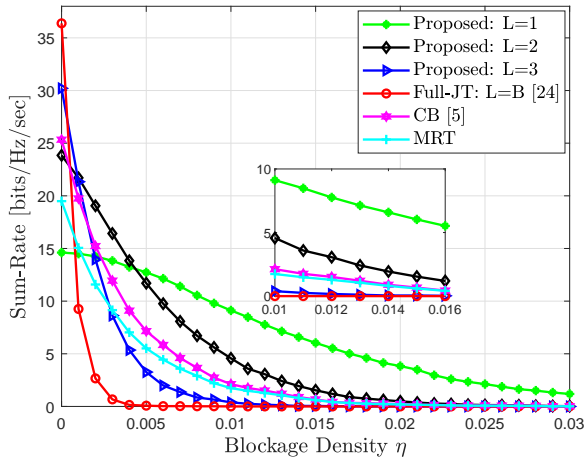}
\caption{Sum-rate as a function increasing blockage density.}
\label{fig:throughput-Results}
\end{figure}
%%%%%%
%%\end{minipage}
%%\hspace{1mm}
%%\begin{minipage}[c]{0.49\textwidth}
\begin{figure}[t]
\setlength\abovecaptionskip{-0.25\baselineskip}
\centering
\setlength\abovecaptionskip{-0.25\baselineskip}
\includegraphics[trim=0.25cm 0.08cm 0.25cm 0.2cm, clip, width=0.95\linewidth]{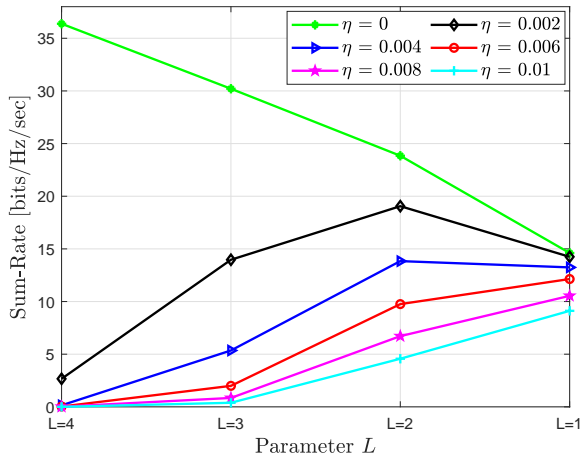}
\caption{Sum-rate as a function different choice of parameter~$L$.}
\label{fig:throughput-L-Res}
%%\end{minipage} 
\end{figure}

Fig.~\ref{fig:throughput-Results} illustrates the trade-off between achievable sum-rate and outage performance 
%%%%with identical user-specific weights (i.e., $\delta_k=1 \ \forall k$) 
for \ac{WSRM} problem. The effective sum-rate $T_\mathrm{e}$ is defined as 
\begin{equation}
\label{eq:outage-def-effective}
T_{\mathrm{e}} \triangleq \big(1- P_{\mathrm{out}}\big)R,
\end{equation}
where $R = \sum_{k\in \mathcal{K}} \log_2 \big(1+\tilde{\gamma}_k\big)$, i.e., when each active user successfully receives the transmit data. 
{It can be observed that when the blockage is not present (i.e., $\eta=0$), with the smaller subset size~$L$, the sum-rate is significantly reduced due to overly a pessimistic estimate of the aggregated SINR~(see Section~\ref{subsec:blockage}). However, it remains relatively more stable even at much higher link blockage density due to improved communication reliability (see Fig.~\ref{fig:outage-Results}). {On the other hand}, with the conventional \ac{JT}, the sum-rate quickly approaches zero with the slight increase in blockage density, due to higher outage. Clearly, there is a trade-off between sum-rate and outage performance, e.g., the outage performance can be improved at the small cost of achievable sum-rate.}
Specifically, for a given outage threshold, we can guarantee a minimum achievable sum-rate and vice versa. In addition, parameter $L$ {(or $\{\mathcal{B}_k, L_k\}_{k\in\mathcal{K}}$ in general)} can be considered as an optimization or selection variable which maximizes the sum-rate for a given blockage density and outage performance, as shown in Fig.~\ref{fig:throughput-L-Res}.
The proposed methods provide robust and resilient connectivity under uncertainties of mmWave radio channel and random blockages, whereas, with the conventional full-\ac{JT} ($L=B$), \ac{CB} and \ac{MRT} methods, sum-rate rapidly decrease towards zero, even if the blockages are slightly~increased.

%% ----- %%%% ----- %%
%\vspace{-10px}
\subsection{Impact of Initialization and Step Sizes}
\label{subsec:Initialization}

%
%
\begin{comment}
 \begin{figure}[t]
%% \setlength\abovecaptionskip{-0.25\baselineskip}
 \centering
 \includegraphics[trim=0.25cm 0.08cm 0.25cm 0.2cm, clip, width=0.5\linewidth]{figs/DifferentStepSize.eps}
 \caption{Convergence performance for Algorithm~$2$ with $L=3$ and $\eta=0$ for different subgradient~($\beta$) and best response~($\psi$) step sizes.}
 \label{fig:Step-Size}
 \end{figure}
%
%
%
%
 \begin{figure}[t]
%% \setlength\abovecaptionskip{-0.25\baselineskip}
 \centering
 \includegraphics[trim=0.25cm 0.08cm 0.25cm 0.2cm, clip, width=0.5\linewidth]{figs/Convergance2.eps}
 \caption{Convergence performance with $L=3$, $\psi = 0.05$, $\beta=0.005$ and $\eta=0$ for Algorithm~$1$~(solid line), Algorithm~$2$ with MRT initialization~(dashed line) and Algorithm~$2$ with random initialization~(dash-dotted line).}
 \label{fig:Convergance}
 \end{figure}
\end{comment}
%
%
%
%
\begin{figure}[t]
%%\begin{minipage}[c]{0.49\textwidth}
\setlength\abovecaptionskip{-0.25\baselineskip}
 \centering
 \setlength\abovecaptionskip{-0.25\baselineskip}
 \includegraphics[trim=0.25cm 0.08cm 0.25cm 0.2cm, clip, width=0.95\linewidth]{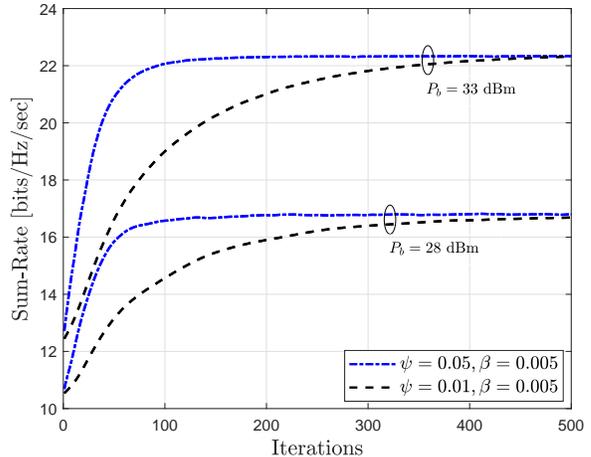}
 \caption{Convergence performance for Algorithm~$2$ with $L=3$ for different subgradient~($\beta$) and best response~($\psi$) step sizes.}
 \label{fig:Step-Size}
 \end{figure}
 %%%%%%%%%%%%%%%%%
%%\end{minipage}
%%\hspace{1mm}
%%\begin{minipage}[c]{0.49\textwidth}
\begin{figure}[t]
\setlength\abovecaptionskip{-0.25\baselineskip}
 \centering
 \setlength\abovecaptionskip{-0.25\baselineskip}
 \includegraphics[trim=0.25cm 0.08cm 0.25cm 0.2cm, clip, width=0.95\linewidth]{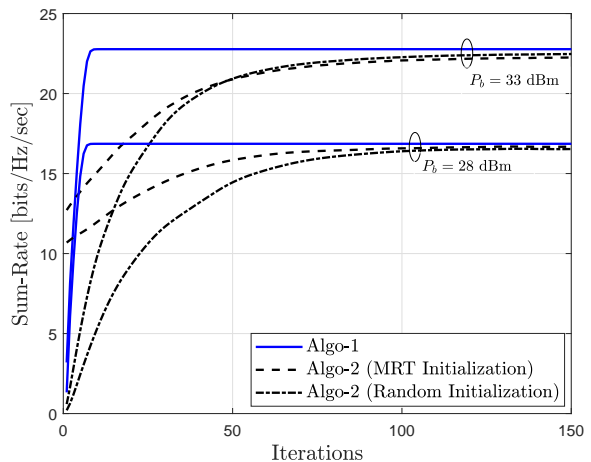}
 \caption{Convergence performance for Algorithm~$1$ and Algorithm~$2$ with $L=3$, $\psi = 0.05$ and $\beta=0.005$.}
 \label{fig:Convergance}
%%\end{minipage} 
\end{figure}
 \begin{figure}[t]
\setlength\abovecaptionskip{-0.25\baselineskip}
 \centering
 \includegraphics[trim=0.25cm 0.08cm 0.25cm 0.2cm, clip, width=0.95\linewidth]{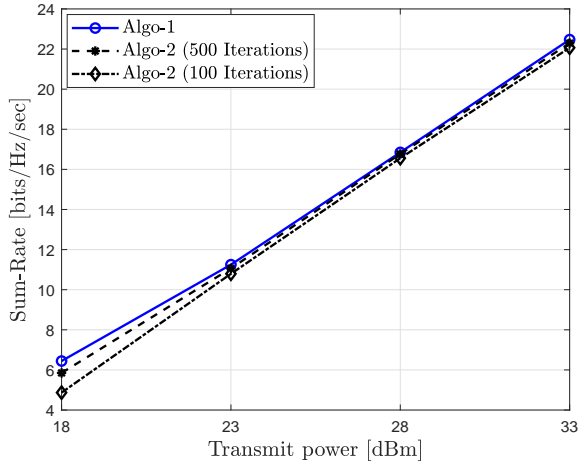}
 \caption{Sum-rate as a function of increasing transmit power with $L=3$ for Algorithm~$1$ and Algorithm~$2$.}
 \label{fig:SNR-Rate}
 \end{figure}
First, in Fig.~\ref{fig:Step-Size} and Fig.~\ref{fig:Convergance}, we examine the convergence performance of Algorithm~\ref{algSCA} and Algorithm~\ref{algKKT}.  For simplicity but without loss of generality, we consider \ac{JT}-\ac{CoMP} scenario with full-coordination i.e., $|\mathcal{B}|=4 $. We set parameter $L=3$ and blockage density $\eta=0$. It can be concluded that convergence is very sensitive to the choice of the step sizes~$(\psi, \beta)$. For example, with the larger value of step sizes, the algorithm can converge with a fewer number of iterations, but might result in more fluctuations to the objective. It should be noted that convergence may not necessarily be monotonic,  which is an inherent feature of subgradient updates (see~\eqref{eq: KKT-i ak}). %~\cite{boyd2003subgradient}.    

%It can be concluded that 
In addition, ingenious choice of the feasible initial points $\big\{\mathbf{{f}}_{b,k}^{\scriptsize (0)}, \tilde{\gamma}_k^{\scriptsize (0)}, a_{k,c}^{\scriptsize (0)} \big\}$ also impact the rate of convergence. For the considered scenario with $N_t \geq K$, a simple \ac{MRT} based initialization for~$\mathbf{{f}}_{b,k}^{\scriptsize (0)} \ \forall (b,k)$ significantly improves the convergence rate and attains near optimal solution with fewer number of iterations, as shown in Fig.~\ref{fig:Convergance}. 
However, irrespective of the choice of initialization point and step sizes, both algorithms tends to converge to the  {local} optimal solution, {on the average}, provided a sufficient number of iterations.

Finally, we compare the sum-rate performance of the low-complexity method based on iteratively solving a set of \ac{KKT} conditions with the solution obtained directly by the optimization toolbox~\cite{cvx}.
Moreover, with the assumption of full-CSI, ({on average}) the achievable performance approaches to the theoretical upper bound. It can be concluded, from Fig.~\ref{fig:SNR-Rate}, that Algorithm~\ref{algKKT} achieves near-optimal performance and the resulting gap in the sum-rate performance is mainly due to insufficient convergence because of the fixed number of maximum iterations. Therefore, the proposed  \ac{KKT} based iterative method provides a low-complexity solution for practical implementations without any significant degradation in the achievable system performance.

%%\vspace{-10px}
\subsection{Hybrid Analog-Digital Beamforming Implementation}
\label{subsec:Hybrid}
While the problem formulation and proposed solutions are generic, they can be easily extended to any multi-point configuration and dense deployments. Until now, we have restricted ourselves to the case where each antenna is equipped with a dedicated RF chain and data converter that enables fully digital signal processing. In this subsection, we provide an implementation for two-stage hybrid analog-digital architecture with coarse-level analog beamforming with a limited number of RF circuits followed by less-complex digital precoding in the digital baseband domain. 

Generally, due to the high power consumption and cost of mixed-signal components in the mmWave frequencies,  analog beamforming is performed using a network of phase-shifters~\cite{Alkhateeb-LimitedFeedback-2015,Yu-Alternating-2016}. To this end, one of the common solutions in the literature is to select the analog beams from a fixed predefined codebook~\cite{Alkhateeb-LimitedFeedback-2015}. We assume that analog beamforming vector $\mathbf{w}_{b,k}$ between a RRU-user pair~($b,k$) is obtained from a fixed beam steering codebook~$\mathcal{W}$ with cardinality $|\mathcal{W}| = 32$. Furthermore, we assume that each RRU~$b$ independently decides  analog beamformers which maximize their local signal power, i.e., based on the following criterion: 
\begin{equation}
\begin{aligned}
\label{eq:AnalogBeamSelection}
& \underset{\mathbf{v}_m \in \mathcal{W}}{\mathrm{maximize}} 
& & \mathbf{w}_{b,k} = \bigl| \mathbf{h}_{b,k}\herm \mathbf{v}_m \bigr|^2
\end{aligned}
\end{equation}

\textit{Case-1:}
For example, let $N_{RF}=K$ be the number of RF circuits at each RRU~$b$, then, from expression~\eqref{eq:AnalogBeamSelection} we obtain $\mathbf{W}_b = [\mathbf{w}_{b,1}, \mathbf{w}_{b,2}, \ldots \mathbf{w}_{b,K}] \in \mathbb{C}^{N_t \text{x} K}$.   
After fixing the user-specific analog beams, each RRU~$b$ estimates the effective channel, i.e., $\widetilde{\mathbf{H}}_b~=~{\mathbf{H}}_b\herm \mathbf{W}_b$ for all $b\in\mathcal{B}$ and then computes the robust digital precoder, as explained in Section~\ref{sec:Precoder-Design}.

\textit{Case-2:}
For example, if we consider $N_{RF}=1$ and $K=4$ then each RRU~$b$ will have at most one active analog beam in a given direction. Therefore, aligning such a directional beam towards a specific user will degrade the achievable SNR for all other active users. However, to efficiently utilize the \ac{JT}-\ac{CoMP} gain, one needs to provide a comparable SNR to all the users. {To do that, we first obtain a compromise transmit beam by appropriate phase-shifts and amplitude scaling to each antenna element using variable gain amplifiers~\cite{Roh-Hybrid-AnalogScaling, Onggosanusi-Modular-2018}, e.g.,  by superimposing best beam of each user~$k$, as $\widetilde{\mathbf{w}}_b=\sum_k \mathbf{w}_{b,k} \big/ \| \sum_k \mathbf{w}_{b,k} \|$ for all $b\in\mathcal{B}$.} It should be noted, in general, $\widetilde{\mathbf{w}}_b \in \mathbb{C}^{N_t \text{x} 1}$ may not satisfy the uni-modulus constraints on beamforming coefficients~\cite{Yu-Alternating-2016, kumar2018reliable}. {Optimization of the compromise multicast analog beam with $N_{RF} < K$ and uni-modulus constraints is an interesting topic for future studies.}
%
%%%%%%%%%%%%%%%%However, compromise receive beam can be optimized with uni-modulus constraints by using the Kronecker decomposition~\cite{Zhu_Kronecker_2017}, which is left for the future work.      
%
After fixing the compromise transmit beam, each RRU estimates the effective channel, i.e., $\widetilde{\mathbf{h}}_b={\mathbf{H}}_b\herm \widetilde{\mathbf{w}}_b \ \forall b\in\mathcal{B}$ and then obtain the digital precoder, as explained in Section~\ref{sec:Precoder-Design}.
 \begin{figure}[t]
 \setlength\abovecaptionskip{-0.25\baselineskip}
 \centering
 \includegraphics[trim=0.25cm 0.08cm 0.25cm 0.2cm, clip, width=0.95\linewidth]{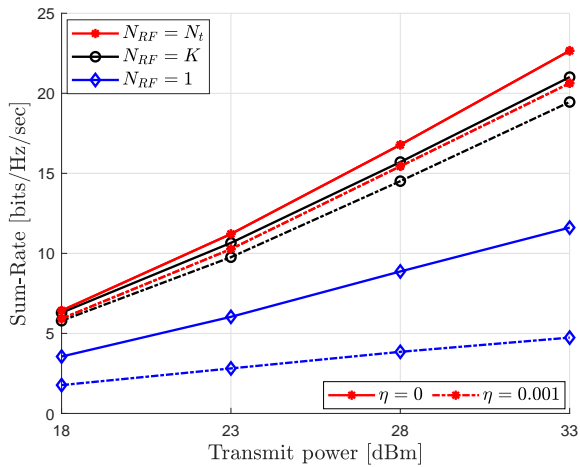}
 \caption{Sum-rate as a function of increasing transmit power for different hybrid analog-digital beamforming configuration with $L=3$ and blockage density $\eta=0$~(solid line), $\eta=0.001$~(dash-dotted line).}
 \label{fig:Hybrid}
 \end{figure}
{Fig.~\ref{fig:Hybrid} illustrate the  sum-rate performance with JT-CoMP scenario, e.g.,  $|\mathcal{B}|=4$. For simplicity but without loss of generality, we set parameter $L=3$ and blockage density $\eta=\{0, \ 0.001\}$}. It can be seen that achievable sum-rate with the two-stage hybrid beamforming architecture is in general lower than full-digital beamforming. This is mainly due to dimensionality reduction of the digital precoder brought by fixed analog beamformers in the two-stage hybrid architecture. 
When $N_{RF}=K$, each RRU implements user-specific analog beam selection and achieves comparable performance to the full-digital beamforming.
However, when $N_{RF}=1$, each RRU implements a compromise transmit beam which is aligned to all $K$ users, thus significantly reducing the achievable analog beamforming gain because of relatively wide analog beams. In addition, the overall system is \ac{DoF} limited in the digital domain, i.e.,  $L < K$, which leads to a significant decrease in the sum-rate performance, specially in the high-SNR conditions.  

In the hybrid architecture, analog beamformers are obtained from a predefined beamforming codebook~\eqref{eq:AnalogBeamSelection}. Thus, the computational complexity of the proposed methods mostly stems from the computation of digital precoder. It can be seen from expression~\eqref{eq:KKT-i iteration} that computations are dominated by matrix multiplications and inversion operations in~\eqref{eq:KKT-i, f*}. Hence, the computational complexity mainly depends on the dimension of the matrix and it is of $\mathcal{O}(N_t^3)$ for inverse operation~\cite{boyd2004convex}. Therefore, hybrid analog-digital beamforming architecture provides dimensionality reductions for digital precoder from $N_t$ to $N_{RF}$. Thus, the computational complexity of the matrix inverse operation is significantly reduced. As an example, the hybrid architecture provides a complexity reduction for the matrix inversion by $\{98.44\%, \ 99.98\%\}$ for $N_t = 16$ and $N_{RF}=\{K, \ 1\}$, respectively. Thus, hybrid architecture achieves the better balance between complexity and system performance.

%
%%%%%%%%%%%%In other words, the hybrid architecture achieves better balance between the computational complexity and achievable system performance by splitting of the signal processing operations in to two stages: analog and digital precoding.

%\vspace{-10px}
%% CONCLUSIONS -------------------------------------------
\section{Conclusion}
\label{sec:conclusion}
In this paper, we studied the trade-off between achievable rate and reliability in mmWave access by exploiting the multi-antenna spatial diversity and CoMP connectivity.  
To combat unpredictable random blockages, a pessimistic estimate of the user-specific rate is obtained over the link blockage combinations, thus providing greatly improved communication reliability.
We devised a low complexity robust beamforming scheme, which is tractable for the practical implementations, based on the best response and subgradient methods, wherein, each RRU specific beamformers are optimized in parallel. Our proposed methods provide a significant reduction in the computational complexity with respect to joint optimization overall RRUs beamforming vectors.
Thus, the proposed methods are scalable to any arbitrary multi-point configuration and dense deployments. 
%
%%%%%For the robust beamformer design, a pessimistic estimate of the user-specific rates is obtained over the link blockage combinations, thus providing greatly improved communication reliability. 
%
Specifically, we proposed a  computationally efficient iterative algorithm for the {WSRM} problem, based on the {SCA} framework and parallelization of the corresponding {KKT} solutions, while accounting for the uncertainties of mmWave radio channel in terms of random link blockages. 
Simulation results manifested the robustness of proposed beamformer design in the presence of random blockages. 
The outage performance and achievable throughput with the proposed methods significantly outperform the baseline scenarios and results in more stable and resilient connectivity for highly reliable {mmWave} communication.
%

% \vspace{-2px}
%% Acknowledgements ---------------------------------------
% \section*{Acknowledgments}
% The research leading to these results has received funding from the Academy of Finland projects: 
% Positioning-aided Reliably-connected Industrial Systems with Mobile mmWave Access (PRISMA) and 6Genesis Flagship (Grant No. 318927). 

%%%%%%%%%%%%%%%%%%%%%%%%%%%%%%%%%%%%%%%%%%%%%%%%%%%%%%%%%%%%%%%%%%%%%%%%%%%%%%%%%%%%%%%%%%%%
%\vspace{-5px}
\appendices
%\vspace{-5px}
\section{}
\label{app:KKT-Conditions}
Considering the Lagrangian in~\eqref{eq:Lagrangian-details}, the stationary conditions are obtained by differentiating~\eqref{eq:Lagrangian-details} with respect to  associated primal optimization variables  $\tilde{\gamma}_k$ and $\overline{\mathbf{f}}_{k}$. 
After some algebraic manipulations, the stationary conditions are given as 
\begin{subequations}
\label{eq:KKT_Derivate}
\begin{align}
\label{eq:Derivate_gamma}
&   \nabla_{\tilde{\gamma}_k} : \sum\limits_{c=1}^{C(L_k)} a_{k,c} \frac{ \sigma_k^2 + \sum_{j\in \mathcal{K}} \bigl| \overline{\mathbf{h}}_{k}\cherm \overline{\mathbf{f}}_{j}\iLoop\bigr| ^2}{(1 + \tilde{\gamma}_k\iLoop)^2} = \frac{\delta_k}{1+\tilde{\gamma}_k} , \\
&  \nabla_{\overline{\mathbf{f}}_{k}} : \overline{\mathbf{f}}_{k}\herm = \bigg( \sum\limits_{b \in \mathcal{B}} z_b \mathbf{E}_b + \sum\limits_{u \in \mathcal{K} \setminus  k} \sum\limits_{c=1}^{C(L_u)}  a_{u,c} \overline{\mathbf{h}}_{u}\comb \overline{\mathbf{h}}_{u}\cherm  \bigg)^{-1} \times \nonumber \\ & \qquad \qquad \qquad \qquad
\Bigg\{ \sum\limits_{j\in\mathcal{K}}
\sum\limits_{c=1}^{C(L_j)} a_{j,c} \frac{\overline{\mathbf{f}}_{k}\iherm \overline{\mathbf{h}}_{j}\comb \overline{\mathbf{h}}_{j}\cherm  }{1 + \tilde{\gamma}_j\iLoop} \Bigg\} , \label{eq:Derivate_CompletePrecoder}
\end{align}
where $\mathbf{E}_b \triangleq \text{diag}\big\{\mathbf{0}, \dots, \mathbb{I}_{N_t} \big({\mathcal{B}(b)}\big), \dots \mathbf{0}\big\}$ is a block diagonal matrix, with all entries are zeros except $\mathbb{I}_{N_t}$ for $b$th RRU. 
From~\eqref{eq:Derivate_CompletePrecoder}, it can be noticed that the computational complexity scales exponentially with the length of joint beamformers~($BN_t$), mainly due to matrix inversion. Furthermore, all coupled and interdependent dual variables $z_b \ \forall b$ must be found simultaneously, which hinders the use of closed-form expressions. Thus, the complexity of algorithm may become intractable in practice, in particular, for dense deployments with large $N_t$ and $B$.    
Here, instead, we implement a parallel optimization framework using the best response approach, which efficiently parallelize the beamformer updates across the distributed RRU with significantly reduced complexity~as
\begin{align}
 \nabla_{\mathbf{f}_{b,k}} : & \  \mathbf{f}_{b,k}\herm = \Big( \mathbb{I}z_b + \sum\limits_{u \in \mathcal{K} \setminus  k} \sum\limits_{c=1}^{C(L_u)} a_{u,c} \mathbf{h}_{b, u}\comb \mathbf{h}_{b, u}\cherm  \Big)^{-1} \nonumber \times \\ & \qquad \qquad 
\Bigg\{ \sum\limits_{j\in\mathcal{K}} \sum\limits_{c=1}^{C(L_j)} a_{j,c} \frac{\big(\overline{\mathbf{f}}_{k}\iherm \overline{\mathbf{h}}_{j}\comb \big) }{1 + \tilde{\gamma}_j\iLoop} \mathbf{h}_{b,j}\cherm \label{eq:Derivate_Precoder} \\ & \qquad \quad \quad
- \sum\limits_{u \in \mathcal{K} \setminus  k } \sum\limits_{c=1}^{C(L_u)} a_{u,c} \bigg( \sum_{g \in \mathcal{B} \setminus  b } \mathbf{f}_{g,k}\herm \mathbf{h}_{g,u}\comb \mathbf{h}_{b,u}\cherm \bigg) \Bigg\}. \nonumber
\end{align}
\end{subequations}
\noindent where $\mathbf{h}_{b,k}\comb = \mathbbm{1}_{\mathcal{G}_k\comb}(b)\mathbf{h}_{b,k} \ \forall (b,k,c)$.
In addition to~\eqref{eq:KKT_Derivate} and the primal-dual feasibility constraints, the \ac{KKT}~conditions also include the complementary slackness conditions as given by
%
%\begin{subequations}
%\label{eq:complementry_slackness}
\begin{flalign}
{a}_{k,c} & \Big\{ I_k({\mathcal{B}}_k\comb) - \mathcal{F}_k\iLoop \big(c, \overline{\mathbf{f}}_{k}, \tilde{\gamma}_k; \overline{\mathbf{f}}_{k}\iLoop, \tilde{\gamma}_k\iLoop \big)  \Big\} = 0 \ \ \forall (k,c), %\ {a}_{k,c} \geq 0 
\label{eq:slackness_akc}
\\
z_b  &  \Big\{ \textstyle \sum_{k \in \mathcal{K}}\|\mathbf{f}_{b,k}\|^2 - P_b \Big\} = 0  \ \ \forall b. %\ z_b \geq 0 
\label{eq:slackness_cb}
\end{flalign}
%\end{subequations}
%
Lets assume the user-specific priority weights $\delta_k > 0 \ \forall k$ (and $\tilde{\gamma}_k \geq 0 \ \forall k$), then from expression~\eqref{eq:Derivate_gamma}, we can observe
\begin{equation}
\label{eq:aK_sum-nonZero}
     \sum\limits_{c=1}^{C(L_k)} a_{k,c} \frac{ \sigma_k^2 + \sum_{j\in \mathcal{K}} \bigr| \overline{\mathbf{h}}_{k}\cherm \overline{\mathbf{f}}_{j}\iLoop\bigr| ^2}{(1 + \tilde{\gamma}_k\iLoop)^2} = \frac{\delta_k}{1+\tilde{\gamma}_k}  > 0 \ \ \forall k.
\end{equation}
Thus, we can infer that at least one of dual-variables~$a_{k,c} \ \forall c$ for each user~$k$ is strictly positive and LHS of \eqref{eq:aK_sum-nonZero} is zero if and only if $\delta_k = 0 \ \forall k$. For simplicity,~\eqref{eq:aK_sum-nonZero} can be rewritten~as 
\begin{equation}
\label{eq:gammaCompute}
   \tilde{\gamma}_k ={\delta_k} \Bigg\{ \sum_{c=1}^{C(L_k)} a_{k,c} \frac{ \sigma_k^2 + \sum_{j\in \mathcal{K}} \bigl| \overline{\mathbf{h}}_{k}\cherm \overline{\mathbf{f}}_{j}\iLoop\bigr| ^2}{(1 + \tilde{\gamma}_k\iLoop)^2} \Bigg\}^{-1}  - 1  \ \ \forall k.  
\end{equation}
The dual-variables $a_{k,c} \ \forall (k,c)$ are coupled and interdependent due to the common SINR constraint, as also seen from~\eqref{eq:KKT_Derivate}~and~\eqref{eq:slackness_akc}. 
Therefore, it is difficult calculate the exact values of these variables in closed-form expressions.
However, all the coupled non-negative Lagrangian multipliers $a_{k,c} \ \forall (k,c)$ can be iteratively solved using the subgradient method, such as based on constrained ellipsoid method. For %SCA
iteration~$i$, the update for the dual variable $a_{k,c}$ with a small positive step size $\beta>0$ can be formulated as 
\begin{equation}
\label{eq:subGrad for aK}
    a_{k,c}\iLoop = \Big( a_{k,c}\xiLoop + \beta \big[ \tilde{\gamma}_k\iLoop - \Gamma_k\iLoop({\mathcal{B}}_k\comb) \big] \Big)^{+} \ \ \forall (k,c) .
\end{equation}
The dual variables $\mathbf{a}^{\scriptsize (0)}=[a_{1,1}^{\scriptsize (0)}, a_{2,1}^{\scriptsize (0)}, \dots, a_{K,C(L_K)}^{(\scriptsize 0)}]\tran$ are initialized with small positive  values. 
\noindent From~\eqref{eq:Derivate_Precoder}, we obtain the transmit~beamformer~as in expression~\eqref{eq:KKT-i, f*}. 
Finally, the dual variables $z_b \ \forall b$ are chosen to satisfy the total power constraints~\eqref{eq:P1-KKT-Pt-2}, using the bisection search method. 

%\vspace{-5px}
\section{}
%\vspace{-5px}
\label{app:success}
The success probability of $k$th user in expression~\eqref{eq:SuccessProb} can be upper bounded by using the binomial theorem, and defined~as 
\begin{equation}
    \label{eq:Success-UpperBound}
    {p}_{k} = \big( |\mathcal{B}_k|- \Psi_k\big)\binom{|\mathcal{B}_k|}{\Psi_k} \int_{t=0}^{1-\overline{q}_{k}}  \! t^{|\mathcal{B}_k|-\Psi_k-1}(1-t)^{\Psi_k} dt,
\end{equation}
where $\Psi_k = |\mathcal{B}_k|-L_k \ \forall k\in\mathcal{K}$. In expression~\eqref{eq:Success-UpperBound}, $\overline{q}_{k}$ is mean blocking probability of each user~$k$ and expressed as $\overline{q}_{k}~=~\frac{1}{|\mathcal{B}_k|}\sum_{b\in\mathcal{B}_k} \widetilde{q}_{b,k}$, where
\begin{equation*}
    \widetilde{q}_{b,k} = \frac{1}{2x_k} \frac{1}{2y_k} \int_{-x_k}^{x_k} \int_{-y_k}^{y_k}  \! 1 - e^{-\eta d_{b,k}} dx_k dy_k \ \ \forall k,
\end{equation*}
%
%%%%and $d_{b,k} = \sqrt{(x_b-x_k)^2+(y_b-y_k)^2}$. We have use $(x_k, y_k)$ and $(x_b, y_b)$ to denotes coordinates of $k$th user and $b$th RRU, respectively. Thus, success probability is obtained by integrating with users location.
%
$d_{b,k} = \sqrt{(x_b-x_k)^2+(y_b-y_k)^2}$ and $(x,y)$ denotes the coordinates in 2D plane. Thus, success probability can be obtained by integrating with respect to users location.

%%%%%$(x, y)$ are the user coordinates and $d_{b,k} = \sqrt{(x_b-x_k)^2+(y_b-y_k)^2}$. Thus, success probability is obtained by integrating with respect to users location.

%%%%\vspace{-15px}
%% REFERENCES --------------------------------------------
\bibliographystyle{IEEEtran}
\bibliography{IEEEabrv,ref_conf_short,ref_jour_short,referencesJournal}

\begin{IEEEbiography}[{\includegraphics[trim=0.058cm 0.058cm 0.058cm 0.058cm, clip, width=1in,height=1.25in,clip,keepaspectratio]{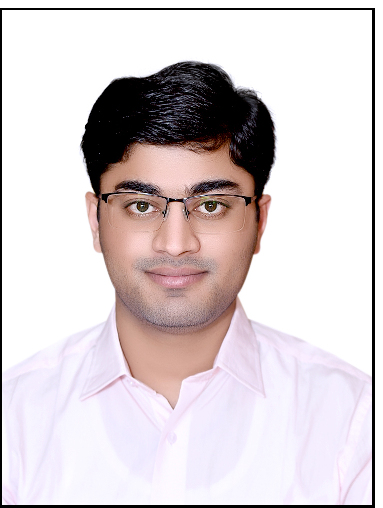}}]{Dileep Kumar}
received the master's degree in communication engineering from the Indian Institute of Technology Bombay (IIT Bombay), India, in 2015. He is currently pursuing the Ph.D. degree at the University of Oulu, Finland. From 2015 to 2017, he worked for {NEC} Corporation, Tokyo, Japan, as a Research Engineer. In 2018, he joined the Centre for Wireless Communications (CWC), University of Oulu.  His research interest includes signal processing for wireless communication systems.
\end{IEEEbiography}

\begin{IEEEbiography}[{\includegraphics[width=1in,height=1.25in,clip,keepaspectratio]{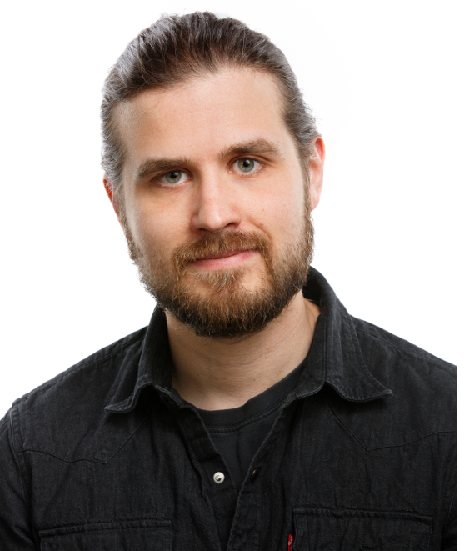}}]{Jarkko Kaleva}(S'11-M'18) received his Dr.Sc. (Tech.) degree in communications engineering from University of Oulu, Oulu, Finland in 2018 with distinction. In 2010, he joined Centre for Wireless Communications (CWC) at University of Oulu, Finland. He is a co-founder of Solmu Technologies, where he is working as the chief software architect. His main research interests are in nonlinear programming, dynamic systems and deep learning.
\end{IEEEbiography}

\begin{IEEEbiography}[{\includegraphics[width=1in,height=1.25in,clip,keepaspectratio]{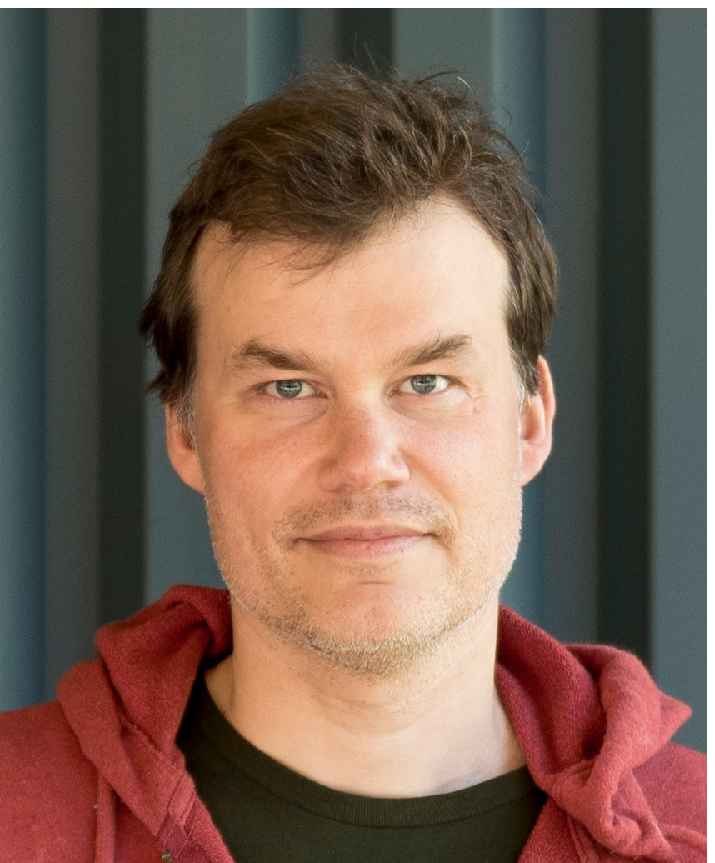}}]{Antti T\"{o}lli}(M'08, SM'14) is an Associate Professor with the Centre for Wireless Communications (CWC), University of Oulu. He received the Dr.Sc. (Tech.) degree in electrical engineering from the University of Oulu, Oulu, Finland, in 2008. From 1998 to 2003, he worked at Nokia Networks as a Research Engineer and Project Manager both in Finland and Spain. In May 2014, he was granted a five year (2014-2019) Academy Research Fellow post by the Academy of Finland. During the academic year 2015-2016, he visited at EURECOM, Sophia Antipolis, France, while from August 2018 till June 2019 he was visiting at the University of California Santa Barbara, USA. He has authored numerous papers in peer-reviewed international journals and conferences and several patents all in the area of signal processing and wireless communications. His research interests include radio resource management and transceiver design for broadband wireless communications with a special emphasis on distributed interference management in heterogeneous wireless networks. He is currently serving as an Associate Editor for IEEE Transactions on Signal Processing.
\end{IEEEbiography}

\end{document}